	\documentclass[10pt,final,doubleside]{IEEEtran}
\normalsize
\usepackage{pifont,bm,multicol,amsfonts,amsmath,color,amssymb,graphicx,mathrsfs,tabularx, epsfig,cite,psfrag,subfigure,algorithm,enumerate,stfloats,algorithm,algorithmic,epstopdf,balance}

\newtheorem{remark}{\underline{Remark}}

\begin{document}
\title{STAR-RIS Aided Integrated Sensing and Communication over High Mobility Scenario
}
\author{
Muye Li, Shun Zhang, \emph{Senior Member, IEEE}, Yao Ge, \emph{Member, IEEE}, \\ Zan Li, \emph{Senior Member, IEEE}, Feifei Gao, \emph{Fellow, IEEE}, Pingzhi Fan, \emph{Fellow, IEEE}

\thanks{
Manuscript received 26 September, 2024; revised 2 February, 2024, accepted 17 March, 2024; date of current version 17 March, 2024.
The work of M. Li and S. Zhang was supported by the National Natural Science Foundation of China under Grant (62271373).
The work of Y. Ge was supported by the RIE2020 Industry Alignment Fund-Industry Collaboration Projects (IAF-ICP) Funding Initiative, as well as cash and
in-kind contribution from the industry partner(s).
The work of P. Fan was supported by NSFC project No.U23A20274.
\emph{(Corresponding author: Shun Zhang.)}
%The work of P. Fan was supported by NSFC Project (No.62020106001).
%\emph{(Corresponding author: Shun Zhang.)}
%This work was supported in part by the National Natural Science Foundation of China under Grant 61871455, 61931017, 6183101, 61901329, in part by the China Postdoctoral Science Foundation under Grant 2019M653557, in part by NSFC Project (No.61731017) and National Key R$\&$D Project (No.2018YFB1801104), in part by the Fundamental Research Funds for the Central Universities and the Innovation Fund of Xidian University under Grant 20109205456, also in part by the SAIC Science and Technology Foundation under Grant No. 1911.
}

\thanks{Muye Li, Shun Zhang, and Zan Li are with the State Key Laboratory of Integrated Services Networks, Xidian University, Xi'an 710071, China (e-mail: myli$\_$96@stu.xidian.edu.cn; zhangshunsdu@xidian.edu.cn, zanli@xidian.edu.cn).}

\thanks{Yao Ge is with the Continental-NTU Corporate Laboratory, Nanyang Technological University, Singapore 637553 (e-mail: yao.ge@ntu.edu.sg).}

\thanks{Feifei Gao is with the Department of Automation, the State Key Laboratory of Intelligent Technologies and Systems, and the State Key for Information Science and Technology (TNList), Tsinghua University, Beijing 100084, China (e-mail: feifeigao@ieee.org).}

\thanks{Pingzhi Fan is with the Key Laboratory of Information Coding and Transmission, CSNMT Int Coop. Res. Centre, Southwest Jiaotong University, Chengdu, 611756, China (e-mail: p.fan@ieee.org).}

}

\maketitle
\vspace{-3mm}
\begin{abstract}

Integrated sensing and communication (ISAC) has become a promising technology for future communication system.
In this paper, we consider a millimeter wave system over high mobility scenario, and propose a novel simultaneous transmission and reflection reconfigurable intelligent surface (STAR-RIS) aided ISAC scheme.
To improve the communication service of the in-vehicle user equipment (UE) and simultaneously track and sense the vehicle with the help of nearby roadside units (RSUs), a STAR-RIS is equipped on the outside surface of the vehicle.
Firstly, an efficient transmission structure for the ISAC scheme is developed, where a number of training sequences with orthogonal precoders and combiners are respectively utilized at BS and RSUs for channel parameter extraction.
Then, the near-field static channel model between the STAR-RIS and in-vehicle UE as well as the far-field time-frequency selective BS-RIS-RSUs channel model are characterized.
By utilizing the multidimensional orthogonal matching pursuit (MOMP) algorithm, the cascaded channel parameters (i.e., the delays, the Doppler frequency shifts, the angles of arrivals, and the angles of departure of the scattering paths) of the BS-RIS-RSUs links can be obtained at the RSUs.
Thus, the vehicle localization and its velocity measurement can be acquired by jointly utilizing these extracted cascaded channel parameters of all RSUs.
Note that the MOMP algorithm can be further utilized to extract the channel parameters of the BS-RIS-UE link for communication service.
With the help of sensing results, the reflection and refraction phase shifts of the STAR-RIS are delicately designed, which can significantly improve the received signal strength for both the RSUs and the in-vehicle UE, and can finally enhance the sensing and communication performance.
Moreover, the trade-off design for sensing and communication is proposed by optimizing the energy splitting factors of the STAR-RIS.
Finally, simulation results are provided to validate the feasibility and effectiveness of our proposed STAR-RIS aided ISAC scheme.

%In this paper, we consider the mmwave system over high mobility scenario, and propose an ISAC scheme for a mobile vehicle and its interior user, where a simultaneous transmission and reflection reconfigurable intelligent surface (STAR-RIS) is equipped on the surface of the vehicle.
%Firstly, the transmission structure for ISAC is designed.
%Then, the far-field time-frequency selective channel model outside the vehicle and the near-field static channel model inside the vehicle are both studied.
%By utilizing the multidimensional orthogonal matching pursuit algorithm, we obtain the cascaded channel parameters from the base station to the roadside units (RSUs), i.e., the delays, the Doppler frequency shifts, the angles of arrivals (AOAs), and the angles of departure (AODs) of each scattering paths.
%Instead of decoupling the parameters of each side of the STAR-RIS, we directly realize the vehicle localization and velocity measurement by gathering the cascaded parameters about all nearby RSUs.
%Moreover, we design and predict the reflection and refraction phase shifts of the STAR-RIS with the help of sensing results.
%Furthermore, by allocating the energy splitting factors, we propose the trade-off design for the performance of sensing and communication.
%Finally, simulation results are provided to validate the feasibility and effectiveness of our proposed STAR-RIS aided ISAC scheme.

\end{abstract}

\maketitle
\thispagestyle{empty}


\section{Introduction}

With the rapid growth of data traffic and the limitation of spectrum resources, novel communication paradigms become necessary to meet the upcoming requirements \cite{background1,background2}.
One possible way is to implement wireless communication networks on higher frequency bands \cite{mmwave1}, such as the millimeter wave (mmWave) and teraherz (THz) bands.
However, a large portion of spectral resources in those bands has been allocated to radar systems \cite{mmwave2}.
Thus, the coexistence of wireless communication networks and radar systems becomes important and challenging.
In recent years, integrated sensing and communication (ISAC) has become a hot research spot for the next generation mobile communication system \cite{ISAC_Mag}.
In ISAC system, communication signal and sensing signal will share same spectral resource, which can significantly improve the spectral utilization.
Compared with the scheme of interference management \cite{interf_manage}, ISAC makes it possible to design merged applications of sensing and communication \cite{ISAC1}, which can further enable the mutual improvement of both functionalities \cite{ISAC2}.
Therefore, ISAC has attracted great attentions of worldwide researchers from both academia and industry.

The key of ISAC lies on the similar channel characteristics of both sensing and communication.
In other words, the recognition of propagation environment is very important for ISAC.
However, with the influence of dynamic scatterers, the propagation environment are always uncertain, which makes the ISAC performance not-guaranteed \cite{RIS_ISAC1}.
Fortunately, the recently proposed reconfigurable intelligent surface (RIS) can control the impinging electromagnetic waves by controlling the reflection coefficients of its reflecting elements \cite{RuiZhang_Mag}.
This enables the adjustment of physical propagation environment.
Even if two communication terminals are blocked by physical obstacles, RIS can help build a virtual line-of-sight (LOS) link to address the blockage problem in the mmWave/THz system \cite{AIRIS}.
Thus, the utilization of RIS can significantly enlarge the coverage and improve the performance of ISAC system.
A number of wireless applications of RIS are summarized in \cite{RIS_survey_re}, including RIS assisted communications in high-frequency bands, RIS-empowered ISAC, and space-air-ground-ocean communications with RIS.

However, traditional RIS can only reflect the incident signal.
Thus, it can only serve terminals at the same side of the RIS.
As a result, if the terminals are located on both sides of the RIS, the benefit of conventional RIS can not be fully exploited.
To tackle this issue, a novel concept of simultaneously transmitting and reflecting RIS (STAR-RIS) has been proposed in \cite{STARRIS_Mag}.
Different from conventional RIS, STAR-RIS can not only reflect but also refract the incident signal, leading to a full-space coverage.
Thus, STAR-RIS has better adaptability of communication environment than conventional RIS, and will have wider application potential for ISAC.

Moreover, ISAC is also very important to high mobility scenario \cite{Mobility1, Mobility2}, where the channel will suffer from large-scale Doppler frequency shift, and its characteristic will change very fast \cite{PDMA1, PDMA2}.
Besides, for in-vehicle terminals, the communication links are usually blocked by the
vehicle body, which can severely harm the signal strength and further influence the communication quality.
In order to maintain the communication performance under such scenario, STAR-RIS aided ISAC can be a promising solution, and it is necessary to develop efficient transmission structure and the target tracking/prediction schemes for performance improvement.

\subsection{Prior Works}

\subsubsection{Literatures on ISAC}

Motivated by the advantage of ISAC, extensive research efforts have been devoted to explore the coexistence of sensing and communication.
In \cite{ISAC_C1}, Li \emph{et. al.} proposed a novel ISAC transmission framework based on the spatially spread orthogonal time frequency space (OTFS) modulation, and designed symbol-wise precoding scheme for communication based on parameters estimated from radar sensing.
In \cite{ISAC_C2}, Yang \emph{et. al.} developed a hybrid simultaneous localization and mapping mechanism that combines active and passive sensing, in which the two sensing modes were mutually enhanced in communication systems.
In \cite{ISAC_C3}, Tong \emph{et. al.} proposed an iterative and incremental joint multi-user communication and environment sensing scheme by exploiting the sparsity of both the structured user signals and the unstructured environment.
In \cite{ISAC_C4}, Xiao \emph{et. al.} proposed a novel full-duplex ISAC scheme that utilizes the waiting time of conventional pulsed radars to transmit communication signals.
%In \cite{ISAC_C5}, Mao \emph{et. al.} considered the challenges in mmWave and low-THz scale, and constructed a data-embedded multi-subband quasi-perfect (MS-QP) waveform through time-domain extension of the designed MS-QP sequence.

\subsubsection{Literatures on RIS aided ISAC}

Since the utilization of RIS can provide an artificially controllable scattering propagation environment, a number of works on RIS aided ISAC emerges.
In \cite{RISISAC_C1}, Zhang proposed to maximize the communication data rate and the mutual information for sensing by jointly optimizing the beamformer at the base station (BS) and the phase shifts at the RIS.
In \cite{RISISAC_C2}, Wang \emph{et. al.} proposed an alternating optimization algorithm to mitigate multi-user interference in RIS aided ISAC system.
In \cite{RISISAC_C3}, Salem \emph{et. al.} proposed to use an active RIS to maximize the achievable secrecy rate of the ISAC system under minimum radar detection signal-to-noise ratio (SNR) and power budget constraints.
In \cite{RISISAC_C4}, Jiang \emph{et. al.} considered RIS aided near-field communication, and proposed the maximum likelihood method and the focal scanning method to sense the location of the receiver.
%In \cite{RISISAC_C5}, Wang \emph{et. al.} derived theoretical limits on the localization and communication performance, where both the continuous intelligent surface and the discrete intelligent surface were considered.
In \cite{RISISAC_C6}, Keykhosravi \emph{et. al.} proposed a joint localization and synchronization algorithm for RIS aided SISO system, where the radial velocities was also estimated.

\subsubsection{Literatures on STAR-RIS aided ISAC}
There are two kinds of realizations for STAR-RIS by using metasurfaces \cite{Prototype1, Prototype2}.
In \cite{Prototype1}, each element of STAR-RIS was composed of a parallel resonant inductance-capacitance tank and small metallic loops to provide the required surface characteristic.
On the other hand, in \cite{Prototype2}, intelligent omni-surface (IOS) was proposed, whose elements are operated by controling the state of positive intrinsic negative diodes, and their phase shifts for transmission and reflection are identical.
For the first kind of realization,
%Wu \emph{et. al.} \cite{STARRIS_C1} proposed a channel estimation scheme that estimate the channel of two users with optimized training (transmission/reflection) pattern.
%In \cite{STARRIS_C2}, Zhu \emph{et. al.} incorporate the concept of index modulation (IM) into STAR-RIS aided non-orthogonal multiple access (NOMA) system to improve the spectral efficiency.
Niu \emph{et. al.} \cite{STARRIS_C3} considers the energy splitting based STAR-RIS aided MIMO system, and  proposed a sub-optimal block coordinate descent algorithm to design the precoding matrices as well as the STAR-RIS coefficients.
In \cite{STARRISISAC_C1}, Wang \emph{et. al.} proposed a STAR-RIS enabled ISAC framework, where the STAR-RIS was utilized to partition the entire space into a sensing space and a communication space.
Besides, for the second kind of realization,
%Zhang \cite{OmniRIS_C1} formulate a joint IOS phase shift design and BS beamforming optimization problem to maximize the received power of multiple mobile users on both sides of the IOS.
%Zhang \emph{et. al.} \cite{OmniRIS_C2} proposed an IOS aided indoor communication system, and embedded an IOS in a wall between two independent access points (APs) to suppress inter-cell interference.
Wu \emph{et. al.} \cite{OmniRIS_C3} proposed an bilayer-IOS (BIOS) structure to achieve flexible reflection and refraction beamforming, and resorted to weighted mean square error minimization approach to enhance the system spectral efficiency.
In \cite{STARRISISAC_C2}, Zhang \emph{et. al.} proposed to realize seamless 360-degree ISAC coverage by collaboratively realizing the joint active and passive beamforming of the dual-function BS and IOS.

\subsection{Motivations and Contributions}
The above works on ISAC mainly focus on the waveform design and localization.
However, velocity is also an important characteristic that should be measured in high mobility scenarios.
Although \cite{RISISAC_C6} estimated the radial velocities along two different directions, the real velocity vector was still not able to be recovered.
In this work, we propose a novel STAR-RIS aided ISAC scheme, where a STAR-RIS is equipped on the outside surface of a vehicle to improve the communication service of the in-vehicle user equipment (UE) and simultaneously reflect signals to the nearby roadside units (RSUs) for tracking the location and velocity of the vehicle.
Specifically, we develop an efficient transmission structure for the ISAC scheme, where a number of training sequences with orthogonal precoders and combiners are respectively utilized at BS and RSUs for parameter extraction.
We also characterize the near-field static channel model of the RIS-UE link, as well as the far-field time-frequency selective channel model of the BS-RIS-RSUs links.
The cascaded channel parameters (i.e., the delays, the Doppler frequency shifts, the angles of arrivals (AOAs), and the angles of departure (AODs) of the scattering paths) of the BS-RIS-RSUs links are obtained at the RSUs by the multidimensional orthogonal matching pursuit (MOMP) algorithm.
With these extracted cascaded channel parameters, the RSUs can jointly realize the vehicle localization and velocity measurement.
Meanwhile, the reflection and refraction phase shifts of STAR-RIS can be designed and predicted with the sensing results.
Finally, we propose a trade-off design for the performance of sensing and communication by optimizing the energy splitting factors of the STAR-RIS.

\subsection{Organizations and Notations}

The rest of this paper is organized as follows.
Section II illustrates the STAR-RIS aided mmWave ISAC system model over high mobility scenario.
Then, the MOMP based channel parameter extraction is proposed in Section III.
Section IV introduces the proposed vehicle sensing scheme, including the vehicle localization and velocity measurement.
Finally, we propose a trade-off design for sensing and communication.
Simulation results are provided in Section V, and conclusions are drawn in Section VI.

Notations: Denote lowercase (uppercase) boldface as vector (matrix).
$(\cdot )^H $, $(\cdot )^T $, $(\cdot )^{*} $, and $(\cdot )^{\dagger} $ represent the Hermitian, transpose, conjugate, and pseudo-inverse, respectively.
$\mathbf I_N $ is an $N \times N $ identity matrix.
$\mathbb E \{\cdot \} $ is the expectation operator.
Denote $|\cdot | $ as the amplitude of a complex value.
$[\mathbf A]_{i,j} $ and $\mathbf A_{\mathcal Q,:}$ (or $\mathbf A_{:, \mathcal Q} $) represent the $(i,j) $-th entry of $\mathbf A $ and the submatrix of $\mathbf A $ which contains the rows (or columns) with the index set $\mathcal Q $, respectively.
$\mathbf x_{\mathcal Q} $ is the subvector of $\mathbf x $ built by the index set $\mathcal Q $.
%$\lfloor p \rfloor $ denotes the largest integer less than or equal to $p $.
%${\boldsymbol\Xi}^{(l-1)} \setminus {\boldsymbol \alpha}^{(l-1)}$ denotes the set ${\boldsymbol\Xi}^{(l-1)}$ expect the element ${\boldsymbol \alpha}^{(l-1)}$.
%The real component of $x $ is expressed as $\Re \{x\}$.
$\text{diag} (\mathbf x)$ is a diagonal matrix whose diagonal elements are formed with the elements of $\mathbf x $.
$(x)_n$ is the mod operation of $x$ with respect to $n$.

\section{STAR-RIS aided MmWave ISAC System Model}

\begin{figure}[htbp]
 \centering
 \includegraphics[width=80mm]{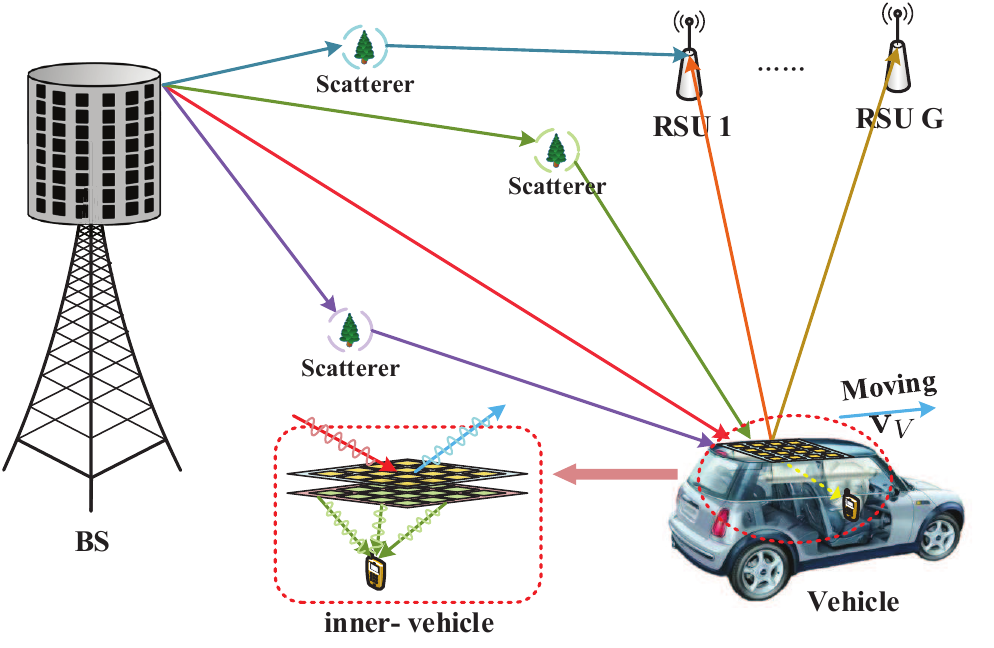}
 \caption{Illustration of STAR-RIS aided downlink ISAC scenario.}
 \label{scene2}
\end{figure}

As shown in \figurename{ \ref{scene2}}, we consider a mmwave communication scenario consisting of one BS and a single antenna UE inside a moving vehicle, where a STAR-RIS is equipped on the outside surface of the vehicle.
Actually, there may be multiple UEs in the vehicle, and we only consider one UE in this paper for simplicity.
The BS is equipped with a $N_B = N_B^x \times N_B^y$ uniform planar array (UPA) and $N_B^{RF}$ radio frequency (RF) chains,
while the STAR-RIS is equipped with an $N_S = N_S^x \times N_S^y$ UPA.
In addition, there are $G$ road side units (RSUs) around the vehicle with $N_R = N_R^x \times N_R^y$ UPAs and $N_R^{RF}$ RF chains.
Note that each RF chain at the BS as well as the RSUs can access to all the antennas by using phase shifters.
Besides, the RSUs are served as passive radars and do not transmit any signals.
Moreover, the sensing results will be gathered at the BS by feedback links, and the BS will further refine the results and optimize the BS and STAR-RIS beamforming designs.
For simplicity, we assume that all the planar arrays are square rather than rectangular.
Since the in-vehicle UE is very close to the STAR-RIS, the channel between the UE and the STAR-RIS should be modelled as near-field channel.
Moreover, the channel from the STAR-RIS to the BS and the RSUs are assumed to be far-field ones.

\subsection{Transmission Structure}

\begin{figure}[htbp]
 \centering
 \includegraphics[width=80mm]{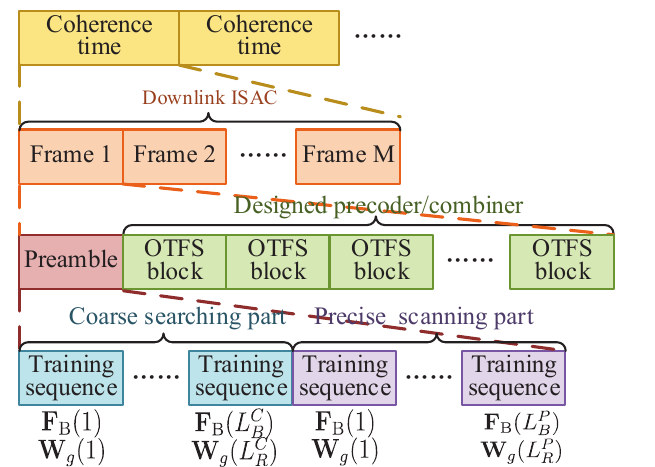}
 \caption{The proposed transmission structure for STAR-RIS aided ISAC.}
 \label{transmission_structure}
\end{figure}

To integrate sensing and communication over downlink (DL) mobility scenario, we propose a transmission structure as shown in \figurename{ \ref{transmission_structure}}.
%For simplicity, we assume that the $N_B^{RF}=1$, and there is only one input data stream at the BS.
The coherence time is divided into several frames, where each frame consists of a preamble for sensing and a number of orthogonal time frequency space (OTFS) modulated blocks for communication.
The preamble is divided into two parts.
One is for coarse beam searching, while the other is for precise beam scanning.
In the first part, the space domain with respect to BS and RSUs are divided into $L_B^C \ll N_B$ and $L_R^C \ll N_R$ parts, respectively.
Thus, the number of pilot sequences with orthogonal precoders and combiners in the first part is $L_O^C = L_B^C L_R^C$.
In the second part, we focus on the directions where signal is detected in coarse beam searching, and implement narrower beams for beam scanning accuracy.
Thus, the second part of the preamble consists of $L_O^P = L_B^P L_R^P$ orthogonal precoded pilot sequences, where $L_B^P$ and $L_R^P$ are the number of beams transmitted by BS and that received by RSUs in this part, respectively.
By contrast, in one-stage full-space precise beam scanning, the space domain with respect to BS and RSUs will usually be divided into $N_B$ and $N_R$ parts, respectively.
Thus, the required number of training sequences within the preamble is $N_B N_R$.
However, the required number of training sequences using the two-stage structure is $L_B^C L_R^C + L_B^P L_R^P$.
Since $L_B^C$, $L_R^C$, $L_B^P$, and $L_R^P$ are usually very small values with respect to $N_B$ and $N_R$, the two-stage beam scanning strategy can significantly reduce the training overhead and computational complexity compared to the one-stage one.

% Compared to one-stage full-space beam scanning strategy, the proposed two-stage one will significantly reduce the training overhead and computational complexity.

%Note that the DL preamble consists of $L_O$ precoded pilot sequences, where each sequence starts with zero padding.
%Besides, each sequence will be precoded at the BS, and be combined at the RSUs.
%To fully cover the range of beam directions, we use orthogonal pairs of the precoder and combiner for different pilot sequences within the preamble.
%\textcolor[rgb]{1.00,0.00,0.00}{Denote the size of beam scanning codebook for the BS and the RSUs as $L_B$ and $L_R$, respectively.}
%Thus, we have $L_O = L_B L_R$.

%Furthermore, the preamble will be received at both the $G$ RSUs and the UE.
For parameter extraction, we focus on the second part of the preamble received at both $G$ RSUs and the UE.
The RSUs extract parameters of the cascaded BS-RIS-RSUs channel links for the localization and velocity measurement of the vehicle, while the UE estimates the cascaded BS-RIS-UE channel link for further transmission.
With sensing results, the BS then adopts OTFS modulation, and organizes pilot symbols and data symbols over delay-Doppler-angle domain with optimized beamforming.
In the meantime, the STAR-RIS will work in fully refraction mode and be delicately designed, while the UE will demodulate the received OTFS blocks for data detection with the help of the estimated channel.
In this work, we focus on the sensing part of the overall ISAC system.
For more detailed OTFS transmission and detection, one can refer to \cite{OTFS_my}.

%Firstly, the UE sends a short pilot sequence and receives by the BS, and the BS will discover the UE, and estimate the AOAs, Doppler frequency shift, and delays of all the scattering paths.
%Then, the BS sends a precoded downlink (DL) preamble towards the direction of the vehicle.
%Note that the DL preamble consists of a number of precoded pilot sequences.
%For different pilot sequences within the preamble, the reflection phase shift matrix and the refraction one are not only different but also orthogonal.
%Further, the preamble will be received by the $J$ RSUs and the UE.
%The RSUs extract the channel parameters between the vehicle and RSUs for the localization and velocity detection of the UE, while the UE estimate the near-field channel between the ORIS and the UE.
%With the knowledge of sensing results, the BS will then adopt OTFS modulation, and organize pilot symbols and data symbols over delay-Doppler-angle domain signal space.
%During OTFS transmission, the ORIS will work in fully refraction mode and be delicately designed, while the UE will demodulate the received OTFS blocks for data detection.

%\begin{figure*}[htbp]
% \centering
% \includegraphics[width=120mm]{transmission_structure.eps}
% \caption{The proposed ISAC transmission structure.}
% \label{transmission_structure}
%\end{figure*}

\subsection{The Structure of the STAR-RIS}

\begin{figure}[htbp]
 \centering
 \includegraphics[width=80mm]{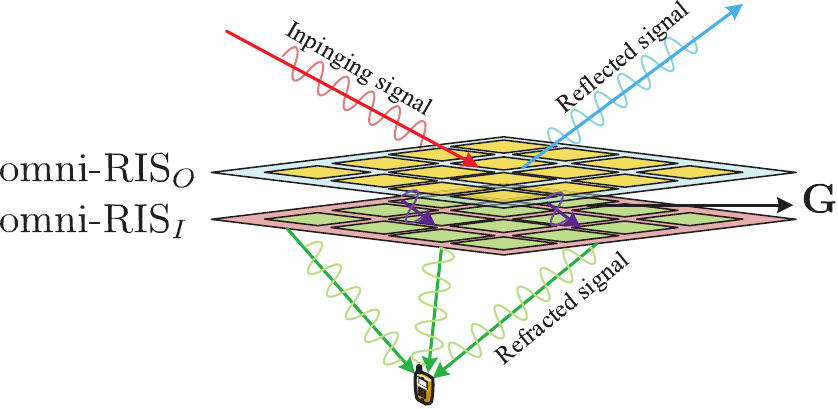}
 \caption{Bilayer structure of the adopted STAR-RIS.}
 \label{ORIS_structure}
\end{figure}

\figurename{ \ref{ORIS_structure}} illustrates the bilayer structure of the utilized STAR-RIS.
The STAR-RIS is composed of two neighboring omni-RISs, where the energy splitting factors of both omni-RISs can be adjusted for different purpose.
For the DL ISAC considered in this paper, the outside one can simultaneously reflect and refract impinging signals to both sides, and the inside one is set to full penetration mode.
Hence, the reflected signal is only related with the outside one, while the refraction signal is first refracted by the outside one, and then penetrate through the inside one.

Define the coefficient matrices of the two omni-RISs as $\mathbf \Omega_O$ and $\mathbf \Omega_I$, respectively.
Then, the overall reflection phase shift matrix $\mathbf \Omega_R$ and transmission phase shift matrix $\mathbf \Omega_T$ can be respectively represented as
$\mathbf \Omega_R = \sqrt{\epsilon_{R}^O} \mathbf \Omega_O $ and
$\mathbf \Omega_T = \sqrt{\epsilon_{T}^O \epsilon_{T}^I} \mathbf \Omega_I \mathbf G \mathbf \Omega_O$, %\begin{align}
%\mathbf \Omega_R = \sqrt{\epsilon_{R}^O} \mathbf \Omega_O,
%\quad\quad
%\mathbf \Omega_T = \sqrt{\epsilon_{T}^O \epsilon_{T}^I} \mathbf \Omega_I \mathbf G \mathbf \Omega_O,
%\end{align}
where $\epsilon_{R}^O$ and $\epsilon_{T}^O$ are respectively the reflection and refraction energy splitting factors of the outside omni-RIS, and $\epsilon_{T}^I = 1$ is the refraction energy splitting factor of the inside omni-RIS.
Note that $\epsilon_{R}^O + \epsilon_{T}^O = 1$ if there is no penetration loss in the omni-RISs \cite{omniRIS1}.
Besides, we define $\boldsymbol \omega_O$ and $\boldsymbol \omega_I$ as the vectors respectively composed by the diagonal elements of $\mathbf \Omega_O$ and $\mathbf \Omega_I$ for further use.
Moreover, $\mathbf G \in \mathbb C^{N_S\times N_S}$ is the channel between the two omni-RISs.
With reference to the channel model in \cite{omniRIS2}, the channel gain from the $n_{s_1}$-th element of one omni-RIS to the $n_{s_2}$-th element of another one can be expressed as
\begin{align}
[\mathbf G]_{n_{s_2},n_{s_1}} \!\!\!=\!\! \sqrt{\frac{a^2}{2\pi d_{n_{s_2},n_{s_1}}^{2}} \frac{z}{d_{n_{s_2},n_{s_1}}}}\!
\exp\!\left(\frac{-\!\jmath 2\pi d_{n_{s_2},n_{s_1}}}{\lambda} \!\right),
\end{align}
where $a$ is the size of scattering elements, $z$ is the distance between the planes of the two omni-RISs, and $d_{n_{s_2},n_{s_1}}$ is the distance between the $n_{s_1}$-th element of one omni-RIS and the $n_{s_2}$-th element of another omni-RIS.
By delicately operating $\boldsymbol \omega_O$, $\boldsymbol \omega_I$, $\epsilon_{R}^O$ and $\epsilon_{T}^O$, we can design optimal $\mathbf \Omega_R$ and $\mathbf \Omega_T$ to achieve balanced performance between sensing and communication.

\subsection{Near-Field Channel Model between STAR-RIS and UE}

To model the near-field channel, we establish a local Cartesian coordinate system $\mathcal C^S$ with respect to the STAR-RIS, and set its origin point at one corner of the STAR-RIS.
Thus, the coordinate of the $(n_S^x, n_S^y)$-th STAR-RIS element can be denoted as $\mathbf p_{n_S^x, n_S^y}^S =  ((n_S^x-1)d, (n_S^y-1)d, 0)^T$, where $d$ is the distance between two adjacent STAR-RIS elements.
Denote the location of the UE within $\mathcal C^S$ as $\mathbf p_U^S = (x_U^S, y_U^S, z_U^S)^T$.
Define $n_s = (n^x_S-1)N^y_S + n^y_S$ as the index of vectorized RIS element, then the distance between the UE and the $n_s$-th STAR-RIS element is
\begin{align}
d_{n_s} =& |\mathbf p_{n_S^x, n_S^y}^S - \mathbf p_U^S|
\notag\\
=& \sqrt{((n_S^x\!-\!1)d \!-\! x_U^S)^2 \!+\! ((n_S^y \!-\! 1)d \!-\! y_U^S)^2 \!+\! (z_U^S)^2}.
\end{align}

Moreover, the quasi-static flat fading near-field channel from the $n_s$-th element of the STAR-RIS to the UE can be represented as
\begin{align}\label{time_do_channel_e}
[\mathbf {h}^{\text{SU}}]_{n_s}(t) = [\mathbf{a}_{\text{SU}}]_{n_s}.
\end{align}
Besides, $[\mathbf{a}_{\text{SU}}]_{n_s}$ denotes the $n_s$-th element of the near-field steering vector of the STAR-RIS under a spherical wavefront,
which can be expressed as
\begin{align}
[\mathbf{a}_{\text{SU}}]_{n_s} = \sqrt{6 \cos^2 (\theta_{n_s})} \frac{e^{\jmath 2\pi \frac{d_{n_s}}{\lambda}}}{4\pi \frac{d_{n_s}}{\lambda}}, n_s = 1, 2, \ldots, N_S,
\end{align}
where the free space path loss model \cite{nearfield1} is adopted. $\theta_{n_s}$ is the elevation AOD from the $n_s$-th element to the user, and $\lambda$ is the wavelength of the carrier.

\subsection{Far-Field Channel Model between BS and RSUs}

It is assumed that there are a number of scatterers in the environment.
Hence, there are multiple scattering paths in the channel between the BS and the STAR-RIS.
In practice, signals in mmWave band fade much faster than those at lower frequency band when propagating and reflecting off a surface \cite{mmwave2}.
Therefore, it is reasonable to assume that all of the NLOS paths only experience single-bounce reflection.
Thus, the time-frequency selective channel of the BS-RIS link can be expressed as
\begin{align}\label{h_BR}
\mathbf {H}^{\text{BS}}_i(t)
=& \sum_{p=1}^{P} h^{\text{BS}}_p e^{\jmath 2 \pi \nu^{\text{BS}}_p (t - \tau^{\text{BS}}_p)} \delta(iT_s-\tau^{\text{BS}}_p)
\notag\\
&\times
\mathbf{a}_{\text{S}} (\theta^{\text{BS}}_{p}, \phi^{\text{BS}}_{p})
\mathbf{a}_{\text{B}}^T (\theta^{\text{B}}_{p}, \phi^{\text{B}}_{p}), i = 0, \ldots, I_{T} - 1,
\end{align}
where $h^{\text{BS}}_p \sim \mathcal {CN}(0, \lambda^{\text{BS}}_p)$ is the channel gain of the $p$-th path between the BS and the STAR-RIS, $\nu^{\text{BS}}_p$ and $\tau^{\text{BS}}_p$ are respectively the Doppler frequency shift and the delay of the $p$-th path.
$i$ is the delay index of the channel, and $I_T$ is the length of the channel over delay domain.
Besides, $\mathbf{a}_{\text{B}} (\theta^{\text{B}}_{p}, \phi^{\text{B}}_{p})$ and $\mathbf{a}_{\text{S}} (\theta^{\text{BS}}_{p}, \phi^{\text{BS}}_{p})$ are the antenna steering vector of the $p$-th path,
where $\{\theta^{\text{B}}_{p}, \phi^{\text{B}}_{p}\}$ and $\{\theta^{\text{BS}}_{p}, \phi^{\text{BS}}_{p}\}$ are respectively the azimuth and elevation AODs of the $p$-th path at the BS and the AOAs of the $p$-th path at the STAR-RIS.
Note that
$
\mathbf{a}_{\text{S}}(\theta, \phi)=\mathbf{a}_{\text{S,x}}(\theta, \phi) \otimes \mathbf{a}_{\text{S,y}}(\theta, \phi)
$,
%\begin{align}
%\mathbf{a}_{\mathbf{R}}(\phi, \psi)=\mathbf{a}_{\text{x}}(\phi, \psi) \otimes \mathbf{a}_{\text{y}}(\phi, \psi),
%\end{align}
where $\otimes$ denotes the Kronecker product,
$
\mathbf{a}_{\text{S,x}}(\theta, \phi)\!=\![1,e^{-\jmath 2\pi\frac{d}{\lambda}\sin(\theta)\sin(\phi)}, \cdots, e^{-\jmath 2\pi(N_S^x-1)\frac{d}{\lambda}\sin(\theta)\sin(\phi)}]^T
$
and
$
\mathbf{a}_{\text{S,y}}(\theta, \!\phi)\!\!=\!\![1,e^{-\!\jmath 2\pi\frac{d}{\lambda}\sin(\!\theta)\!\cos(\!\phi)},\cdots,e^{-\!\jmath 2\pi(N_S^x-\!1)\frac{d}{\lambda}\!\sin(\!\theta)\!\cos(\!\phi)}]^T
$
%\begin{align}
%\mathbf{a}_{\text{R,x}}(\theta, \phi)&=[1,e^{-\jmath 2\pi\frac{d}{\lambda}\sin(\theta)\sin(\phi)}, \cdots, e^{-\jmath 2\pi(N_x-1)\frac{d}{\lambda}\sin(\theta)\sin(\phi)}]^T, \label{xvector}\\
%\mathbf{a}_{\text{R,y}}(\theta, \phi)&=[1,e^{-\jmath 2\pi\frac{d}{\lambda}\sin(\theta)\cos(\phi)},\cdots,e^{-\jmath 2\pi(N_y-1)\frac{d}{\lambda}\sin(\theta)\cos(\phi)}]^T \label{yvector}
%\end{align}
are respectively the array response vectors along the elevation and the azimuth dimensions,
with $d = \frac{\lambda}{2}$ representing the inter-element spacing.
$\mathbf{a}_{\text{B}}$ can be defined in the same way.

The received signal at the $g$-th RSU consists of two parts.
One is the signals reflected by the STAR-RIS, and the other is that scattered directly by the static scatterers in the environment.
The first part of the channel is cascaded by the channel between the BS and the vehicle and that between the vehicle and the $g$-th RSU.
Therefore, it is reasonable to assume that $\{\theta^{\text{B}}_{p}, \phi^{\text{B}}_{p}\}_{p=1}^{P}$ in \eqref{h_BR} are the same with that in the first part of the channel from the BS to the $g$-th RSU.
Since the RSUs are close to the vehicle, the LOS path between the vehicle and the RSUs will occupy dominant power of the channel.
Thus, we assume that the channel between the STAR-RIS and any RSU only consists of the LOS path.
Hence, the time-frequency selective channel from the BS to the $g$-th RSU can be represented as
\begin{align}\label{h_BRS}
\mathbf {H}^{\text{B},g}_i(t)
\!=&\! \sum_{p=1}^{P}\! h^{\text{B},g}_p e^{\jmath 2 \pi \nu^{\text{B},g}_p (t - \tau^{\text{B,g}}_p)} \delta(iT_s \!\!-\!\! \tau^{\text{B},g}_p)
\mathbf{a}_{\text{R}} (\theta^{g}, \!\phi^{g})
\notag\\
&\times
\mathbf{a}_{\text{S}}^T (\theta^{\text{SR}_g}, \!\phi^{\text{SR}_g})
\mathbf \Omega_R (t)
\mathbf{a}_{\text{S}} (\theta^{\text{BS}}_{p}, \!\phi^{\text{BS}}_{p})
\mathbf{a}_{\text{B}}^T (\theta^{\text{B}}_{p}, \!\phi^{\text{B}}_{p})
\notag \\
&+
\!\!\sum_{p'=P+1}^{P_g}\!\! h^{\text{B},g}_{p'} e^{\jmath 2 \pi \nu^{\text{B},g}_{p'} (t - \tau^{\text{B},g}_{p'})} \delta(iT_s \!\!-\!\! \tau^{\text{B},g}_{p'})
\notag\\
&\times\!
\mathbf{a}_{\text{R}} (\theta^{g}_{p'}, \phi^{g}_{p'})
\mathbf{a}_{\text{B}}^T (\theta^{\text{B},g}_{p'}, \phi^{\text{B},g}_{p'})
,
i \!\!=\! 0, \ldots, I_{T} \!-\! 1,
\end{align}
where $h^{\text{B},g}_p \sim \mathcal {CN}(0, \lambda^{\text{B},g}_p)$ is the channel gain of the $p$-th path of the channel, $\nu^{\text{B},g}_p$ and $\tau^{\text{B},g}_p$ are respectively the Doppler frequency shift and the delay of the $p$-th path.
$\mathbf{a}_{\text{R}} (\theta^{g}_p, \phi^{g}_p)$ is the antenna steering vector of the $p$-th path, and $\theta^{g}_p$ and $\phi^{g}_p$ are respectively the azimuth and elevation AOAs of the $p$-th path at the $g$-th RSU.
Besides, $P_g - P$ is the number of scattering paths from the BS to the  $g$-th RSU without passing through the STAR-RIS.
$\{\theta^{\text{SR}_g}, \phi^{\text{SR}_g}\}$ are the elevation and azimuth AODs from the vehicle to the $g$-th RSU.
For $p = 1,2, \ldots, P$, it can be extracted that $\nu^{\text{B},g}_p = \nu^{\text{BS}}_p + \nu^{\text{SR}_g}$ and
$\tau^{\text{B},g}_p = \tau^{\text{BS}}_p + \tau^{\text{SR},g}$,
where $\nu^{\text{SR}_g}$ and $\tau^{\text{SR}_g}$ are the Doppler frequency shift and the delay between the vehicle and the $g$-th RSU, respectively.
%Furthermore, $\mathbf \Omega_R(t) = \alpha_R \text{diag}(\boldsymbol \omega_R(t))$ is the RIS reflection phase shift matrix, where $[\boldsymbol \omega_R (t)]_{n_r} = e^{\jmath 2\pi \beta^T_{n_r}(t)}$.
%For further use, we define the refraction phase shift matrix of the ORIS here as $\mathbf \Omega_T(t) = \alpha_T \text{diag}(\boldsymbol \omega_T(t))$, where $[\boldsymbol \omega_T(t)]_{n_r} = e^{\jmath 2\pi \beta^T_{n_r}(t)}$.

\subsection{Received Signal Model}
Define each training sequence transmitted by the BS as
$\mathbf T = [ \mathbf t_0, \mathbf t_1, \ldots, \mathbf t_{N_T-1}]^T \in \mathbb C^{N_T \times N^{RF}_B}$ with $N_T$ represents the length of the sequence.
Note that we do not adjust the phase shift matrices during the training sequence, we will omit the time index $t$ of $\mathbf \Omega_R$ and $\mathbf \Omega_T$ in the following for notational simplicity.
The received signal for the $n_t$-th time-slot of the $l_O^P$-th training sequence ($l_O^P = L_B^P (l_R^P - 1) + l_B^P$, $l_B^P = 1,\ldots, L_B^P $, $l_R^P = 1,\ldots, L_R^P $) at the $g$-th RSU can be represented as
\begin{align}
\mathbf y^{g}_{n_t, l_o^P}
%=&
%\sum_{p=1}^{P} h^{\text{B},g}_p e^{\jmath 2 \pi \nu^{\text{B},g}_p (((l_o-1)(N_{CP}+N_T) + n_t)T_s - \tau^{\text{B},g}_p)}
%\mathbf W_{g}^H(l_R)
%\mathbf{a}_{\text{RSU}} (\theta^{g}, \phi^{g})
%\mathbf{a}_{\text{R}}^T (\theta^{\text{SR}_g}, \phi^{\text{SR}_g})
%\mathbf \Omega_R
%\notag \\
%&\times
%\mathbf{a}_{\text{R}} (\theta^{\text{BS}}_{p}, \phi^{\text{BS}}_{p})
%\mathbf{a}_{\text{B}}^T (\theta^{\text{B}}_{p}, \phi^{\text{B}}_{p})
%\mathbf F_{\text{B}}(l_B)
%\mathbf t_{(n - \tau^{\text{B},g}_p/T_s)_{N_T}}
%\notag\\
%&+\sum_{p'=P+1}^{P_g} h^{\text{B},g}_{p'} e^{\jmath 2 \pi \nu^{\text{B},g}_{p'} (((n_O-1)(N_{CP}+N_T) + n_t)T_s - \tau^{\text{B},g}_{p'})}
%\mathbf W_{g}^H(l_R)
%\mathbf{a}_{\text{RSU}} (\theta^{g}_{p'}, \phi^{g}_{p'})
%\mathbf{a}_{\text{B}}^T (\theta^{\text{B},g}_{p'}, \phi^{\text{B},g}_{p'})
%\notag\\
%&\times
%\mathbf F_{\text{B}}(l_B)
%\mathbf t_{(n - \tau^{\text{B},g}_{p'}/T_s)_{N_T}}
%+ \mathbf n^g_{n_t,l_o}
%\notag \\
=& \sum_{p=1}^{P_g} \widetilde{h}^{\text{B},g}_p e^{\jmath 2 \pi \nu^{\text{B},g}_p (((l_O^P-1)(N_{CP}+N_T) + n_t)T_s - \tau^{\text{B},g}_p)}
\notag\\
&\times
\mathbf W_{g}^H(l_R^P)
\mathbf{a}_{\text{R}} (\theta^{g}_p, \phi^{g}_p)
\mathbf{a}_{\text{B}}^T (\theta^{\text{B},g}_{p}, \phi^{\text{B},g}_{p})
\notag\\
&\times
\mathbf F_{\text{B}}(l_B^P)
\mathbf t_{(n_t - \tau^{\text{B},g}_p/T_s)_{N_T}} + \mathbf n^g_{n_t,l_O^P},
\label{RSU_rec_1}
\end{align}
where $\widetilde{h}^{\text{B},g}_p = h^{\text{B},g}_p \mathbf{a}_{\text{S}}^T (\theta^{\text{SR}_g}, \phi^{\text{SR}_g})
\mathbf \Omega_R  \mathbf{a}_{\text{S}} (\theta^{\text{BS}}_{p}, \phi^{\text{BS}}_{p})$ and $\{\theta^{g}_p, \phi^{g}_p\} = \{\theta^{g}, \phi^{g}\}$ for $p = 1,2,\ldots, P$.
And $\widetilde{h}^{\text{B},g}_p = h^{\text{B},g}_p$ for $p = P+1, \ldots, P_g$.
Besides, $\mathbf F_{\text{B}}(l_B^P)$ and $\mathbf W_{g}^H(l_R^P)$ are the precoder at the BS and the combiner at the $g$-th RSU of the $l_O^P$-th training sequence, respectively.
$\mathbf n^g_{n_t,l_O^P} \sim \mathcal{CN}(0, \sigma_N^2)$ is the additive white Gaussian noise (AWGN) of the $g$-th RSU at the $n_t$-th time slot of the $l_O^P$-th training sequence.
By defining $\bar{h}^{\text{B},g}_p = \widetilde{h}^{\text{B},g}_p e^{-\jmath 2 \pi \nu^{\text{B},g}_p \tau^{\text{B},g}_p}$,
$\mathbf v_{l_O^P} (\nu^{\text{B},g}_p) = e^{\jmath 2 \pi \nu^{\text{B},g}_p \! ((l_O^P\!-\!1)(\!N_{CP} \!+\! N_T\!)) T_s} \![1, \!e^{\jmath 2 \pi \nu^{\text{B},g}_p T_s}\!, \ldots, e^{\jmath 2 \pi \nu^{\text{B},g}_p (\!N_T\!-\!1)T_s}]^T$,
and $[\mathbf A_d(\tau^{\text{B},g}_p)]_{:,n_t} = [\delta(1-(n_t-\tau^{\text{B},g}_p /T_s)_{N_T}), \delta(2-(n_t-\tau^{\text{B},g}_p /T_s)_{N_T}), \ldots, \delta(N_T-(n_t-\tau^{\text{B},g}_p /T_s)_{N_T})]^T$,
\eqref{RSU_rec_1} can be rewritten as
\begin{align}
\mathbf y^{g}_{n_t, l_O^P}
\!\!=&
\sum_{p=1}^{P_g} \!
\bar{h}^{\text{B},g}_p
[\mathbf v_{l_O^P} (\nu^{\text{B},g}_p)]_{n_t}\!
\mathbf W_{g}^H(l_R^P)
\mathbf{a}_{\text{R}} (\theta^{g}_p, \phi^{g}_p)
\notag\\
&\!\!\times \!\!
\mathbf{a}_{\text{B}}^T (\theta^{\text{B},g}_{p}, \phi^{\text{B},g}_{p})
\mathbf F_{\text{B}}(l_B^P)
\mathbf T^T \!
[\mathbf A_d(\tau^{\text{B},g}_p)]_{:,n_t}
\!\!\!+\!\!
\mathbf n^g_{n_t,l_O^P},
\label{RSU_rec_2}
\end{align}
and the received signal $\mathbf Y^{g}_{l_O^P}$ at the $g$-th RSU by stacking the whole $l_O^P$-th pilot sequence can be further derived as
\begin{align}
\mathbf Y^{g}_{l_O^P}
\!=&\!
\sum_{p=1}^{P_g}\!
\bar{h}^{\text{B},g}_p
\mathbf W_{g}^H(l_R^P)
\mathbf{a}_{\text{R}} (\theta^{g}_p, \!\phi^{g}_p)
\mathbf{a}_{\text{B}}^T (\theta^{\text{B},g}_{p}, \!\phi^{\text{B},g}_{p})
\mathbf F_{\text{B}}(l_B^P)
\notag\\
&\times
\mathbf T^T\!
\mathbf A_d(\tau^{\text{B},g}_p)
\text{diag}(\mathbf v_{l_O^P} (\nu^{\text{B},g}_p))
\!+\!  \mathbf N^g_{l_O^P},
\end{align}
where $\mathbf Y^g_{l_O^P} = [\mathbf y^g_{1,l_O^P}, \ldots, \mathbf y^g_{N_T,l_O^P}] \in \mathbb C^{N_R^{RF} \times N_T}$ and $\mathbf N^g_{l_O^P} = [\mathbf n^g_{1,l_O^P}, \ldots, \mathbf n^g_{N_T,l_O^P}] \in \mathbb C^{N_R^{RF} \times N_T}$.

Meanwhile, the received signal for the $l_O^P$-th training sequence at UE can be written as
\begin{align}
\mathbf y^{\text{UE}}_{l_O^P}
%=&
%\sum_{p=1}^{P}
%h_p^{\text{BS}}
%\mathbf{a}_{\text{UR}}^T
%\mathbf \Omega_T
%\mathbf{a}_{\text{R}} (\theta^{\text{BS}}_{p}, \phi^{\text{BS}}_{p})
%\mathbf{a}_{\text{B}}^T (\theta^{\text{B}}_{p}, \phi^{\text{B}}_{p})
%\mathbf F_{\text{B}}(l_B)
%\mathbf T^T
%\mathbf A_d(\tau^{\text{B},g}_p)
%\text{diag}(\mathbf v_{l_o} (\nu^{\text{BS}}_p))
%+ \mathbf n^{\text{UE}}(l_o)
%\notag\\
=&
\sum_{p=1}^{P}
\bar{h}^{\text{BU}}_p
\mathbf{a}_{\text{B}}^T (\theta^{\text{B}}_{p}, \phi^{\text{B}}_{p})
\mathbf F_{\text{B}}(l_B^P)
\notag\\
&\times
\mathbf T^T
\mathbf A_d(\tau^{\text{BS}}_p)
\text{diag}(\mathbf v_{l_O^P} (\nu^{\text{BS}}_p))
+ \mathbf n^{\text{UE}}_{l_O^P},
\label{UE_rec_1}
\end{align}
where $\bar{h}^{\text{BU}}_p = h_p^{\text{BS}}
\mathbf{a}_{\text{SU}}^T
\mathbf \Omega_T
\mathbf{a}_{\text{S}} (\theta^{\text{BS}}_{p}, \phi^{\text{BS}}_{p})$, and $\mathbf n^{\text{UE}}_{l_O^P}$ is the AWGN of the $l_O^P$-th sequence at the UE.

\begin{remark}
Exactly, there must exist some direct links from the BS to the UE.
However, due to the physical blockage of the vehicle body, the power of the electromagnetic waves penetrated into the vehicle will attenuate a lot.
By contrast, without considering the loss caused by the hardware achitecture of the STAR-RIS, it only causes negligible penetration loss when the electromagnetic waves come into the vehicle with the refraction phase shifting operation.
Therefore, the STAR-RIS aided link will contribute the dominant path power of the whole BS-UE channel.
Thus, it can be assumed that there is no other existed direct link from the BS to the UE.
\end{remark}

\section{Channel Parameter Extraction via MOMP}
In the sensing part, we take the $g$-th RSU as an example, and resort to MOMP algorithm for parameter extraction.
Note that the proposed method can be directly applied at other RSUs, and can also be simplified for the UE.
\subsection{Problem Formulation}

Since the channel is modeled as the combination of a number of scattering paths, sparsity appears in the AOA of the RSU, the AOD of the BS, the delay, and Doppler dimensions.
Besides, since orthogonal precoder and combiner are implemented respectively at the BS and the RSUs for different training sequences, the received signal for each training sequence corresponds to different AODs at the BS as well as different AOAs at the RSUs.
Thus, the angular parameters of each training sequence can be determined easily, and we mainly focus on the estimation of other channel parameters.
Specifically, if the precoder and combiner of the relative received sequences do not align with the scattering paths, the signal power will be very low and close to the noise power.
Therefore, the AOAs and AODs can be estimated by searching the index of the received sequences with highest power.
Then, we can focus on estimating the delays and Doppler frequency shifts of the paths.

The choice for the dictionaries of the sets about the channel parameters can be defined as
$
\mathbf \Psi_{\theta,\phi}^{\text{R}} \!=\! [\mathbf a_{\text{R}}(\{\bar{\theta},\bar{\phi}\}_1),
%\mathbf a_{\text{RSU}}(\{\bar{\theta},\bar{\phi}\}_2),
\ldots,
\mathbf a_{\text{R}}(\{\bar{\theta},\bar{\phi}\}_{N_{\theta,\phi}^{\text{R}}})]
$,
$
\mathbf \Psi_{\theta,\phi}^{\text{B}} \!=\! [\mathbf a_{\text{B}}(\{\bar{\theta},\bar{\phi}\}_1),
%\mathbf a_{\text{B}}(\{\bar{\theta},\bar{\phi}\}_2),
\ldots,
\mathbf a_{\text{B}}(\{\bar{\theta},\bar{\phi}\}_{N_{\theta,\phi}^{\text{B}}})]
$,
$
\mathbf \Psi_{\nu}(l_O^P) \!=\! [\mathbf v_{l_O^P}(\bar{\nu}_1),
%\mathbf v_{n_o}(\bar{\nu}_2),
\ldots,
\mathbf v_{l_O^P}(\bar{\nu}_{N_{\nu}})]
$,
and
$
\mathbf \Psi_{\tau} \!=\! [\mathbf a_d(\bar{\tau}_1),
%\mathbf a_d(\bar{\tau}_2),
\ldots,
\mathbf a_d(\bar{\tau}_{N_{\tau}})]
$,
where $[\mathbf a_d(\bar{\tau})]_{n_T} = \delta(n_T - \bar{\tau}/T_s)$
.
%\begin{align}
%\mathbf \Psi_{\theta,\phi}^{g} =& [\mathbf a_{\text{RSU}}(\{\bar{\theta},\bar{\phi}\}_1), \ldots, \mathbf a_{\text{RSU}}(\{\bar{\theta},\bar{\phi}\}_{N_{\theta,\phi}^{g}})],
%\\
%\mathbf \Psi_{\theta,\phi}^{\text{B}} =& [\mathbf a_{\text{B}}(\{\bar{\theta},\bar{\phi}\}_1), \ldots, \mathbf a_B(\{\bar{\theta},\bar{\phi}\}_{N_{\theta,\phi}^{\text{B}}})],
%\\
%\mathbf \Psi_{\nu}(l_o) =& [\mathbf v_{n_o}(\bar{\nu}_1), \ldots, \mathbf v_{n_o}(\bar{\nu}_{N_{\nu}})],
%\\
%\mathbf \Psi_{\tau} =& [\mathbf a_d(\bar{\tau}_1), \ldots, \mathbf a_d(\bar{\tau}_{N_{\tau}})].
%\end{align}
Thus, the number of channel parameters to be searched is $K = 4$.
For notational simplicity, we define $N_1^s = N_{\theta,\phi}^{\text{R}}$, $N_2^s = N_{\theta,\phi}^{\text{B}}$, $N_3^s = N_{\nu}$, $N_4^s = N_{\tau}$ as the dimensions of the above dictionaries, respectively.
Besides, we define the dimension of the elements within each dictionary as $N_1^a = N_R$, $N_2^a = N_B$, $N_3^a = N_{T}$, $N_4^a = N_{T}$.

With this multi-dimensional dictionary configuration, and ignoring quantization effects caused by the finite resolution of the dictionaries, we can define $\mathcal C \in \mathbb C^{N_1^s \times N_2^s \times N_3^s \times N_4^s}$ as
\begin{align}
[\mathcal C]_{\mathbf j}
=
\left\{
\begin{array}{ll}
   \bar{h}^{\text{B},g}_p & \text{if }
       \begin{array}{l}
       \{\theta^{g}_p, \phi^{g}_p \}
       = \{\bar{\theta}, \bar{\phi}\}_{j_1}
       \\
       \{\theta^{\text{B},g}_{p}, \phi^{\text{B},g}_{p}\}
       = \{\bar{\theta}, \bar{\phi}\}_{j_2}
       \\
       \nu^{\text{B},g}_p = \bar{\nu}_{j_3}
       \\
       \tau^{\text{B},g}_p = \bar{\tau}_{j_4}
       \end{array}
    \\
   0 & \text{otherwise}
 \end{array}
 \right\},
\end{align}
where $\mathbf j = (j_1,j_2,j_3,j_4)$ is the set of dictionary indexes.
And we define $\mathcal J = \{\mathbf j \in \mathbb N^{K} \ \text{s.t.} \ j_{k} \leq N^s_{k}, \forall k \leq K\}$ for further use.
Then, the equivalent channel gain representation $\mathcal H_u \in \mathbb C^{N_R \times N_B \times N_V}$ of the $u$-th delay tap over angle-delay-Doppler domain in tensor form can be finally written as
\begin{align}
&[\mathcal H_u(l_O^P)]_{n_r, n_b, n_{\nu}} \notag\\
&\quad=\!\! \sum_{\mathbf j \in \mathcal J}
\![\mathbf \Psi_{\theta,\phi}^{\text{R}}]_{n_r,\!j_1}
\![\mathbf \Psi_{\theta,\phi}^{\text{B}}]_{n_b,\!j_2}
\![\mathbf \Psi_{\nu}(l_O^P)]_{n_{\nu},\!j_3}
\![\mathbf \Psi_{\tau}]_{u,\!j_4}
\![\mathcal C]_{\mathbf j}.
\end{align}

Since our purpose is to transform the channel estimation problem into the multi-dimensional sparse estimation problem,
we can relabel the sub-indexes by their dictionary counterparts for cleaner formulation by substituting the entry indexes $(n_r, n_b, n_{\nu}, u)$ with $\mathbf i = (i_1, i_2, i_3, i_4)$ as
$
[\mathcal H_{i_4}(l_O^P)]_{i_1, i_2, i_3}
=
\sum\limits_{\mathbf j \in \mathcal J} \prod\limits_{k=1}^{K} [\mathbf \Psi_{k}(l_O^P)]_{i_k,j_k} [\mathcal C]_{\mathbf j}
$
%\begin{align}
%[\mathbf H_{i_4}(l_o)]_{i_1, i_2, i_3}
%=&
%%\sum_{\mathbf j \in \mathcal J} [\mathbf \Psi_{\theta,\phi}^{g}]_{i_1,j_1}
%%[\mathbf \Psi_{\theta,\phi}^{\text{B}}]_{i_2,j_2}
%%[\mathbf \Psi_{\nu}(l_o)]_{i_3,j_3}
%%[\mathbf \Psi_{\tau}]_{i_4,j_4} [\mathbf C]_{\mathbf j}
%%\\
%%=&
%\sum_{\mathbf j \in \mathcal J} \prod_{k=1}^{N_P} [\mathbf \Psi_{k}(l_o)]_{i_k,j_k} [\mathbf C]_{\mathbf j}
%\end{align}
for every combination of $i_k \leq N_k^a, \forall k \leq K$, where $\mathbf \Psi_1(l_O^P) = \mathbf \Psi_{\theta,\phi}^{\text{R}}$,
$\mathbf \Psi_2(l_O^P) = \mathbf \Psi_{\theta,\phi}^{\text{B}}$,
$\mathbf \Psi_3(l_O^P) = \mathbf \Psi_{\nu}(l_O^P)$,
$\mathbf \Psi_4(l_O^P) = \mathbf \Psi_{\tau}$.
And we define $\mathcal I = \{\mathbf i \in \mathbb N^{K} \ \text{s.t.} \ i_{k} \leq N^a_{k}, \forall k \leq K\}$ for further use.
This allows us to rewrite the received signal at the $n_r^{RF}$-th RF chain of the $g$-th RSU for the $l_O^P$-th training sequence as
\begin{align}
[\mathbf Y^{g}_{l_O^P}]_{n_r^{RF}\!,:}
\!\!=& \!\!
%\sum_{p=0}^{P_g}
%\bar{h}^{\text{B},j}_p
%[\mathbf W_{g}^H (l_R)]_{m^{RF},:}
%\mathbf{a}_{\text{RSU}} (\theta^{g}_p, \phi^{g}_p)
%\mathbf{a}_{\text{B}}^T (\theta^{\text{B},g}_{p}, \phi^{\text{B},g}_{p})
%\mathbf F_{\text{B}}(l_B)
%\mathbf T^T
%\mathbf A_D(\tau^{\text{B},g}_p)
%\text{diag}(\mathbf v_{l_o} (\nu^{\text{B},g}_p))
%+ \mathbf n
%\\
%=&
\sum_{\mathbf i \in \mathcal I}
[\mathbf W_{g}(l_R^P\!)]_{i_1,n_r^{RF}}^{*}\!
[\mathbf F_{\text{B}}\!(l_B^P\!)]_{i_2,:}\!
\mathbf T^T\!\!
\mathbf A_d(i_4 T_s\!)
\mathbf A_{D}(i_3)
\notag\\
&\times
\!\sum_{\mathbf j \in \mathcal J}\!
\Big( \!\prod_{k=1}^{K} [\mathbf \Psi_{k}\!(l_O^P)]_{i_k,\!j_k} [\mathcal C]_{\mathbf j}\Big) \!+\! [\mathbf N^g_{l_O^P}]_{n_r^{RF}\!,:},
\end{align}
where $\mathbf A_{D}(i_3)$ is an all-zero matrix with only $[\mathbf A_{D}(i_3)]_{i_3,i_3} = 1$.
Then, the weight of the contribution of $\sum\limits_{\mathbf j \in \mathcal J} \prod_{k=1}^{K} [\mathbf \Psi_{k}(l_O^P)]_{i_k,j_k} [\mathcal C]_{\mathbf j}$ to $[\mathbf Y^{g}_{l_O^P}]_{n_r^{RF},n_t}$ is
\begin{align}
[\mathbf \Phi_{n_r^{RF}} (l_O^P)]_{n_t, \mathbf i} =&
[\mathbf W_{g}(l_R^P)]_{i_1,n_r^{RF}}^{*}
[\mathbf F_{\text{B}}(l_B^P) \mathbf T^T]_{i_2, (n_t-i_4)_{N_T}}
\notag\\
&\times
\delta(i_3-n_t)
,
\end{align}
which is denoted as the measurement matrix.
Taking into account that the noise is white, the maximum likelihood estimator is given by the minimum mean square estimator as \eqref{MOMP_problem} at the top of the next page.
\begin{figure*}[htbp]
\begin{align}\label{MOMP_problem}
\arg\min\limits_{\mathcal C} \Big(
\Big\|
%[\mathbf y^{g}_{m^{RF}}(l_o)]^T
[\mathbf Y^{g}_{l_O^P}]_{n_r^{RF},:}^T
- \sum_{\mathbf i} \sum_{\mathbf j}
[\mathbf \Phi_{n_r^{RF}} (l_O^P)]_{:,\mathbf i}
\Big(\prod_{k=1}^{K}[\mathbf \Psi_{k}(l_O^P)]_{i_k,j_k}\Big) [\mathcal C]_{\mathbf j}\Big\|^{2}
\Big),
\end{align}
\hrulefill
\end{figure*}
and it can be solved by the MOMP algorithm for a sparse solution.

\subsection{MOMP based Channel Parameter Extraction}

For the received signal of the $l_O^P$-th sequence at the $n_r^{RF}$-th RF chain which has higher signal power, only the signal of one scattering path is involved.
Thus, we can only search for one set of $\mathbf j$ with each received sequence.
Note that the AODs at the BS and the AOAs at the $g$-th RSU can be derived by
\begin{align}
\{\theta_{l_B^P}^{\text{B},g}, \!\phi_{l_B^P}^{\text{B},g}\} \!\!=&\!
\arg\max\limits_{\{\theta, \phi\}}
\|\mathbf{a}_{\text{B}}^T (\theta, \!\phi)
\mathbf F_{\text{B}}(l_B^P)\|^2,
\\
\{\theta_{l_R^P,n_r^{RF}}^{g}, \!\phi_{l_R^P,n_r^{RF}}^{g}\} \!\!=&\!
\arg\max\limits_{\{\theta, \phi\}}
|\mathbf{a}_{\text{R}}^T (\theta, \!\phi)
[\mathbf W_{g}(l_R^P)]_{:,n_r^{RF}}|,
\end{align}
respectively.
Therefore, the dictionary indexes $j_1$ and $j_2$ for the $l_O^P$-th sequence at the $n_r^{RF}$-th RF chain can be directly determined.
Note that the estimated AODs at the BS are coarse estimates, and can be refined by further beam scanning within more narrow ranges.
Besides, consider the case where the angles are not on the grids, we can gather the received sequences of several neighbor grids with higher signal strength, and select the central one of them to determine the objective received sequence and its angle dictionary indexes.
Then, the remained dictionary indexes $j_3$ and $j_4$ will be both searched to maximize the projection with $[\mathbf Y^{g}_{l_O^P}]_{n_r^{RF},:}$.
In this problem, the projection and matching step is equivalent to the expression
\begin{align}\label{problem1}
\arg \!\!\!\!\!\max\limits_{j_k, k\!=\!\{3,4\}} \!\!\!\left(\!
\frac{
\!\!\Big\|\![\!\mathbf Y^{g}_{l_O^P}]_{n_r^{R\!F}\!\!,:}^{*}\!
\Big(\!\!\sum\limits_{\mathbf i} [\!\mathbf \Phi\!_{n_r^{R\!F}} \!(\!l_O^P)]_{:,\mathbf i}
\!\!\prod\limits_{k=1}^{K} \!\![\!\mathbf \Psi_k \!(\!l_O^P)]_{i_k\!,j_k} \!\!\Big)
\!\Big\|^{2}
}
{
\Big\|\!
\Big(\sum\limits_{\mathbf i} [\mathbf \Phi_{n_r^{RF}}(l_O^P)]_{:,\mathbf i}
\prod\limits_{k=1}^{K}[\mathbf \Psi_k(l_O^P)]_{i_k,j_k}\Big)
\!\Big\|^{2}}
\!\!\right)\!\!.
\end{align}
By defining $\mathbf y_{\mathbf \Phi}({l_O^P, n_r^{RF}}) \in \mathbb C^{1\times N_1^s \times N_2^s \times N_3^s \times N_4^s}$ as $[\mathbf y_{\mathbf \Phi}({l_O^P, n_r^{RF}})]_{1,\mathbf i} = [\mathbf Y^{g}_{l_O^P}]_{n_r^{RF},:}^{*} [\mathbf \Phi_{n_r^{RF}}(l_O^P)]_{:,\mathbf i}$,
the problem \eqref{problem1} can be rewritten as
\begin{align}\label{problem2}
\arg \!\!\!\!\!\!\max\limits_{j_k, k=\{3,4\}} \!\!\!\left(\!
\frac{
\!\Big\|\!
\Big(\!\! \sum\limits_{\mathbf i}
[\mathbf y_{\mathbf \Phi}({l_O^P,n_r^{RF}})]_{1,\mathbf i}
\!\!\prod\limits_{k=1}^{K} \![\mathbf \Psi_k(l_O^P)]_{i_k,j_k} \!\Big)
\!\Big\|^{2}
}
{
\Big\|\!
\Big(\!\sum\limits_{\mathbf i} [\mathbf \Phi_{n_r^{RF}} (l_O^P)]_{:,\mathbf i}
\prod\limits_{k=1}^{K}[\mathbf \Psi_k(l_O^P)]_{i_k,j_k} \!\Big)\!
\Big\|^{2}}
\!\!\right)\!.
\end{align}

Since searching all the parameters of one scattering path simultaneously will incur huge computational complexity, we turn to alternatively search the parameters, and refine them by means of a number of iterations.
For example, in the $l$-th iteration, the refinement of $j_k$ can be implemented by
\eqref{problem3} at the top of the next page.
\begin{figure*}[htbp]
\begin{align}\label{problem3}
\arg \!\!\!\!\max\limits_{j_k, k=\{3,4\}} \left(
\frac{
\Big\|
\Big(\sum\limits_{\mathbf i}
[\mathbf y_{\mathbf \Phi}({l_O^P, n_r^{RF}})]_{1,\mathbf i}
[\mathbf \Psi_k(l_O^P)]_{i_k,j_k}
\prod\limits_{k'\neq k}[\mathbf \Psi_{k'}(l_O^P)]_{i_{k'},\widehat{j}^{(l-1)}_{k'}}\Big)
\Big\|^{2}
}
{
\Big\|
\Big(\sum\limits_{\mathbf i}
[\mathbf \Phi_{n_r^{RF}} (l_O^P)]_{:,\mathbf i}
[\mathbf \Psi_k(l_O^P)]_{i_k,j_k}
\prod\limits_{k'\neq k}[\mathbf \Psi_{k'}(l_O^P)]_{i_{k'},\widehat{j}^{(l-1)}_{k'}}\Big)
\Big\|^{2}}
\right).
\end{align}
\begin{align}\label{problem_init}
\arg \!\!\!\!\max\limits_{j_k, k=\{3,4\}}
\!\!\left(
\frac{
\sum\limits_{k''\in\overline{\mathcal E}}\!
\sum\limits_{i_{k''}=1}^{N_{k''}^s}
\!\Big\|
\sum\limits_{k'''\in{\mathcal E} \cup \{k\}}
\sum\limits_{i_{k'''}=1}^{N_{k'''}^s}
[\mathbf y_{\mathbf \Phi}({l_O^P,n_r^{RF}})]_{1,\mathbf i}
[\mathbf \Psi_k(l_O^P)]_{i_k,j_k}
\prod\limits_{k'\in{\mathcal E}}
[\mathbf \Psi_{k'}(l_O^P)]_{i_{k'},j_{k'}}
\Big\|^{2}\!
}
{
\sum\limits_{k''\in\overline{\mathcal E}}\!
\sum\limits_{i_{k''}=1}^{N_{k''}^s}
\!\Big\|
\sum\limits_{k'''\in{\mathcal E} \cup \{k\}}
\sum\limits_{i_{k'''}=1}^{N_{k'''}^s}
\sum\limits_{\mathbf i} [\mathbf \Phi_{n_r^{RF}} (l_O^P)]_{:,\mathbf i}
[\mathbf \Psi_k(l_O^P)]_{i_k,j_k}
\prod\limits_{k'\in{\mathcal E}}
[\mathbf \Psi_{k'}(l_O^P)]_{i_{k'},j_{k'}}
\Big\|^{2}}\!
\right).
\end{align}
\hrulefill
\end{figure*}

For the first iteration, the parameters are also alternatively initialized.
For the search of initial ${j_k, k=\{3,4\}}$, define $\mathcal E$ as the set of dictionary indexes that we have an initial estimation, and define $\overline{\mathcal E}$ as the set of dictionary indexes that we do not have an initial estimation excluding $k$.
Note that $k \notin \mathcal E$, $k \notin \overline{\mathcal E}$, and $\mathcal E \cup \{k\} \cup \overline{\mathcal E} = \{1,2,\ldots, K\}$.
By splitting the indexes among different sets, and implementing some relaxations \cite{MOMP}, \eqref{problem3} can be further derived as \eqref{problem_init} at the top of the next page.

After searching the dictionary indexes for the parameters of the path, the corresponding equivalent channel gain $[\mathcal C]_{\widehat{\mathbf j}}$ should be calculated.
By utilizing the measurement matrix and the dictionary matrices, we have
\begin{align}\label{calculate_C_hat_j}
[\widehat{\mathcal C}]_{\widehat{\mathbf j}} \!=\!\!
\frac{
\big(\!\sum\limits_{\mathbf i} \![\mathbf \Phi_{n_r^{RF}} (\!l_O^P)]_{:,\mathbf i}
\!\!\prod\limits_{k=1}^{K}
\!\![\mathbf \Psi_k(\!l_O^P)]_{i_k, \widehat{j}_k}
\!\big)\!^H\!
\mathbf [\mathbf y_{\mathbf \Phi}({\!l_O^P,n_r^{R\!F}}\!)]}
{
\left\|\sum\limits_{\mathbf i} \mathbf [\mathbf \Phi_{n_r^{RF}} (l_O^P)]_{:,\mathbf i}
\prod\limits_{k=1}^{K}[\mathbf \Psi_k(l_O^P)]_{i_k, \widehat{j}_k}\right\|^2}.
\end{align}

After the implementation of the MOMP algorithm for all the received sequences with highest signal power, the equivalent channel gain as well as the channel parameters of all the scattering paths can be obtained.
Note that the UE can also acquire its related channel parameters by resorting to the MOMP algorithm in a similar way.

\section{Vehicle Sensing and Trade-Off Design of Sensing and Communication}

\subsection{Recognition of the STAR-RIS Aided Channel at the RSUs}
For user localization and velocity measurement, we should firstly separate the STAR-RIS aided channel from the whole channel between the BS and each RSU.
From \eqref{RSU_rec_1}, one can notice that the cascaded scattering paths reflected by the STAR-RIS contain a certain degree of Doppler frequency shift.
%However, other paths scattered by static scatterers have zero Doppler frequency shift.
Besides, the paths to be recognized have the same AOA at the $g$-th RSU, while others usually have different AOAs.
With those observations, we develop a path recognition algorithm for the RSUs as illustrated in {\bf Algorithm \ref{alg:path_recognition}}.
Firstly, the paths with non-zero Doppler frequency shift are picked out.
Then, the picked paths are divided into different sets, where each set consists of the paths that have the same AOA at the $g$-th RSU.
Finally, due to the reflection of the STAR-RIS, the set which contains maximal number of paths will be chosen.

\begin{algorithm}
	\caption{Path Recognition Algorithm for the $g$-th RSU}
	\label{alg:path_recognition}
	\renewcommand{\arraystretch}{0.5}
	\begin{algorithmic}[1]
        \STATE {\bf input:} $ \{\{\widehat{\theta}_p^{g},
        \widehat{\phi}_p^{g}\},
        \widehat{\nu}_p^{\text{B},g},
        \widehat{\tau}_p^{\text{B},g} \}_{p=1}^{P_g} $.
        \STATE {\bf initialize:} $\mathcal D^{R_g} = \emptyset$,
        $\mathcal D_a^{R_g} = \emptyset $, and $\mathcal A_a^{R_g} = \emptyset$ for $a = 1, 2, \ldots,P_g$.
        \FOR {$p = 1, 2, \ldots, P_g$}
            \IF {$\widehat{\nu}_p^{\text{B},g} \neq 0 $}
                \FOR {$a = 1, 2, \ldots, P_g$}
                    \IF {$\mathcal D_a^{R_g} = \emptyset $ or $\mathcal A_a^{R_g} = \{\widehat{\theta}_p^{g}, \widehat{\phi}_p^{g}\} $}
                        \STATE $\mathcal D_a^{R_g} = \mathcal D_a^{R_g} \cup \{p\}$, $\mathcal A_a^{R_g} = \{\widehat{\theta}_p^{g},
        \widehat{\phi}_p^{g}\}$.
                        \STATE {\bf break.}
                    %\ELSE
%                        \IF {$\mathcal A_a^{R_j} = \{\widehat{\theta}_p^{\text{RSU}_j},
%        \widehat{\phi}_p^{\text{RSU}_j}\} $}
%                            \STATE $\mathcal D_a^{R_j} = \mathcal D_a^{R_j} \cup \{p\}$.
%                            \STATE {\bf break.}
%                        \ENDIF
                    \ENDIF
                \ENDFOR
            \ENDIF
        \ENDFOR
        \STATE $\max = 0$, $a_{\max} = 0$.
        \FOR {$a = 1, 2, \ldots, P_g$}
            \IF {$|\mathcal D_a^{R_g}| \ge \max $}
                \STATE $\max = |\mathcal D_a^{R_g}|$, $a_{\max} = a$.
            \ENDIF
        \ENDFOR
        \STATE {\bf output:} $\mathcal D^{R_g} = \mathcal D_{a_{\max}}^{R_g}$.
	\end{algorithmic}
\end{algorithm}

\subsection{Vehicle Localization}

Since the BS and the RSUs are always pre-deployed, it can be assumed that their locations are known in priori.
In addition, the directions of their antenna planes are also assumed to be known a priori.
To simplify the derivations and design of sensing, we establish a global Cartesian coordinate system $\mathcal C^B$ and fix its origin point at a corner of the BS antenna plane.
Its $x$ and $y$ axes are along the direction of the BS antennas while its $z$ axis can be determined by right-hand rule.
Denote the locations of the BS and RSUs as $\mathbf p_{\text{B}} = [0,0,0]^T$ and $\mathbf p_{g} = [x_g, y_g, z_g]^T$, respectively.
In addition, define the normal vector of the BS antenna plane and RSU antenna planes as $\mathbf n_{\text{B}} = [0,0,1]^T$ and $\mathbf n_{g}$, respectively.
Furthermore, we establish the local coordinate system of the $g$-th RSU as $\mathcal C^{R_g}$ in the way similar with $\mathcal C^B$, in which the location of the BS and the $g$-th RSU can be represented as $\mathbf p_{\text{B}}^{R_g} = [x_B^{R_g},y_B^{R_g},y_B^{R_g}]^T$ and $\mathbf p_{g}^{R_g} = [0,0,0]^T$, respectively.

%\begin{figure*}[htbp]
% \centering
% \includegraphics[width=100mm]{BS_coordinate_system.eps}
% \caption{BS global coordinate system.\textcolor[rgb]{1.00,0.00,0.00}{\bf\{$p_{RIS}$ should be $p_{S}$\}}}
% \label{global_coordinate_system}
%\end{figure*}

After the sub-channel recognition, we have obtained the AOAs at the $g$-th RSU from the STAR-RIS, denoted by $\{{\widehat\theta}^{g}, {\widehat\phi}^{g}\}_{g=1}^{G}$.
Hence, the location of the STAR-RIS must be on the line with the direction $\{{\widehat\theta}^{g}, {\widehat\phi}^{g}\}$ passing through $\mathbf p_{g}$.
In addition, the cascaded delays $\widehat{\tau}_{1}^{\text{B},g}$ of the LOS paths for BS-RIS-RSUs links are also acquired.
Thus, the whole LOS path length is $L_{1}^{\text{B},g} = c \widehat{\tau}_{1}^{\text{B},g}$, where $c$ is the light speed.
With the location of the BS and the $g$-th RSU, one ellipsoid can be established for the LOS reflection path.
With geometric relationships, the location of the STAR-RIS must be at the intersection of the RSU-RIS line and the ellipsoid.
By deriving the locations of the STAR-RIS with respect to $G$ RSUs (i.e., $G$ intersections of the corresponding ellipsoids and lines), the estimated location of the STAR-RIS can be regarded as the center of the $G$ intersection points.

Denote the distance between the BS and the $g$-th RSU as $d^{\text{B},g} = |\mathbf p_{g}|$, which is also the focal length of the ellipsoid.
Then, the length of three semi-major axis related with the $g$-th ellipsoid can be calculated as $a_g = \frac{L_{1}^{\text{B},g}}{2}$ and $b_{g,1} = b_{g,2} = \sqrt{\left(\frac{L_{1}^{\text{B},g}}{2}\right)^2 - \left(\frac{d^{\text{B},g}}{2}\right)^2} \triangleq b_{g}$.

\begin{figure}[htbp]
 \centering
 \includegraphics[width=80mm]{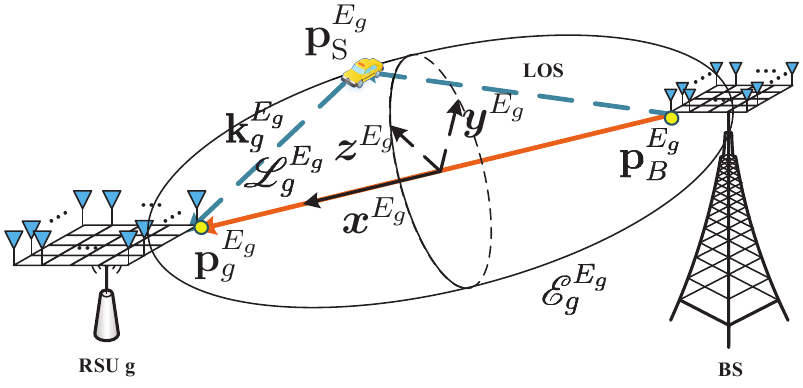}
 \caption{Local coordinate system of the $g$-th Ellipsoid.}
 \label{Ej_local_coordinate_system}
\end{figure}

To simplify the representation of the ellipsoid equation, we establish a local coordinate system $\mathcal C^{E_g}$ as shown in \figurename{ \ref{Ej_local_coordinate_system}}.
Set the direction of $\mathbf p_{g}$ as its $x$ axis, and define the middle point between the BS and the $g$-th RSU as its origin point.
Note that $\mathcal C^{E_g}$ can be derived by parallel moving $\mathcal C^{B}$ along $\mathbf p_{g}$ and then rotating $\mathcal C^{B}$ around its three axes.
Since the lengths of two short radius of the ellipsoid are the same, the rotation angle around the $x$ axis can be set as zero for simplicity in coordinate system transformation.
The ellipsoid equation under $\mathcal C^{E_g}$ can be represented as
\begin{align}
\mathscr E_{g}^{E_g}: \frac{x^2}{a_{g}^2} + \frac{y^2}{b_{g}^2} + \frac{z^2}{b_{g}^2}= 1.
\end{align}

Besides, with the recognized $\{{\widehat\theta}^{g}, {\widehat\phi}^{g}\}$,
the direction vector of the path from the STAR-RIS to the $g$-th RSU can be defined as
$\mathbf k_{g} = [\cos \widehat\theta^{g} \cos\widehat\phi^{g},  \cos \widehat\theta^{g} \sin\widehat\phi^{g}, \sin\widehat\theta^{g}]^T$.
Note that $\mathbf k_{g}$ is the slope of the RSU-RIS line in $\mathcal C^{R_g}$, and should be rotated into $\mathcal C^{E_g}$.
Define the rotation angles from $\mathcal C^{R_g}$ to $\mathcal C^{E_g}$ around the $x$, $y$, and $z$ axes as $\beta_x^{E_g}$, $\beta_y^{E_g}$, and $\beta_z^{E_g}$, respectively.
Then, the corresponding rotation matrices are respectively defined as
\begin{align}
\mathbf R_x^{E_g} =&
\left(\!\!\!
  \begin{array}{ccc}
   \setlength{\arraycolsep}{0.2pt}
    1 & 0 & 0 \\
    0 & \cos\!{\beta_x^{E_g}} & -\!\sin\!{\beta_x^{E_g}} \\
    0 & \sin\!{\beta_x^{E_g}} & \cos\!{\beta_x^{E_g}} \\
  \end{array}
\!\!\!\!\right), \label{rotation_Ejx}
\\
\mathbf R_y^{E_g} =&
\left(\!\!\!\!
  \begin{array}{ccc}
    \cos\!{\beta_y^{E_g}} & 0 & \sin\!{\beta_y^{E_g}} \\
    0 & 1 & 0 \\
    -\!\sin\!{\beta_y^{E_g}} & 0 & \cos\!{\beta_y^{E_g}} \\
  \end{array}
\!\!\!\right), \label{rotation_Ejy}
\\
\mathbf R_z^{E_g} =&
\left(\!\!\!\!
  \begin{array}{ccc}
    \cos\!{\beta_z^{E_g}} & -\!\sin\!{\beta_z^{E_g}} & 0 \\
    \sin\!{\beta_z^{E_g}} & \cos\!{\beta_z^{E_g}} & 0 \\
    0 & 0 & 1 \\
  \end{array}
\!\!\!\right), \label{rotation_Ejz}
\end{align}
%\begin{align}
%\mathbf R_x^{E_g} =&
%\left(
%  \begin{array}{ccc}
%    1 & 0 & 0 \\
%    0 & \cos{\beta_x^{E_g}} & -\sin{\beta_x^{E_g}} \\
%    0 & \sin{\beta_x^{E_g}} & \cos{\beta_x^{E_g}} \\
%  \end{array}
%\right), \label{rotation_x_Ej}
%\\
%\mathbf R_y^{E_g} =&
%\left(
%  \begin{array}{ccc}
%    \cos{\beta_y^{E_g}} & 0 & \sin{\beta_y^{E_g}} \\
%    0 & 1 & 0 \\
%    -\sin{\beta_y^{E_g}} & 0 & \cos{\beta_y^{E_g}} \\
%  \end{array}
%\right), \label{rotation_y_Ej}
%\\
%\mathbf R_z^{E_g} =&
%\left(
%  \begin{array}{ccc}
%    \cos{\beta_z^{E_g}} & -\sin{\beta_z^{E_g}} & 0 \\
%    \sin{\beta_z^{E_g}} & \cos{\beta_z^{E_g}} & 0 \\
%    0 & 0 & 1 \\
%  \end{array}
%\right), \label{rotation_z_Ej}
%\end{align}
where $\beta_x^{E_g} = 0$, and
\begin{align}
\beta_y^{E_g} = \frac{\pi}{2} - \arccos\frac{(\mathbf p_{\text{B}}^{R_g})^T \mathbf e_{z}^{R_g}}{\|\mathbf p_{\text{B}}^{R_g}\| \cdot \|\mathbf e_{z}^{R_g}\|} ,
\\
\beta_z^{E_g} = \arccos \frac{\big(\mathbf p_{\text{B}}^{R_g} - \frac{(\mathbf p_{\text{B}}^{R_g})^T \mathbf e_{z}^{R_g}}{\|\mathbf e_{z}^{R_g}\|} \mathbf e_{z}^{R_g}\big)^T \mathbf e_{x}^{R_g}}
{\big\|\mathbf p_{\text{B}}^{R_g} - \frac{(\mathbf p_{\text{B}}^{R_g})^T \mathbf e_{z}^{R_g}}{\|\mathbf e_{z}^{R_g}\|} \mathbf e_{z}^{R_g}\big\| \cdot \big\|\mathbf e_{x}^{R_g}\big\|}
\end{align}
are the angles of inversely rotation operation around the $x$, $y$ and $z$ axes, respectively.  $\mathbf e_{y}^{R_g}$ and $\mathbf e_{z}^{R_g}$ respectively represent the unit vectors along the $y$ and $z$ axis in $\mathcal C^{R_g}$.
Hence, the direction vector $\mathbf k_{g}$ in the coordinate system $\mathcal C^{E_g}$ can be represented as
$
\mathbf k_{g}^{E_g} = \mathbf R_y^{E_g} \mathbf R_z^{E_g} \mathbf k_{g} /
\|\mathbf R_y^{E_g} \mathbf R_z^{E_g} \mathbf k_{g}\|
$.
%\begin{align}
%\mathbf k_{g}^{E_g} =  \frac{\mathbf R_y^{E_g} \mathbf R_z^{E_g} \mathbf k_{g}}
%{\|\mathbf R_y^{E_g} \mathbf R_z^{E_g} \mathbf k_{g}\|}.
%\end{align}
Then, the RSU-RIS line equation in $\mathcal C^{E_g}$ passing across the STAR-RIS can be represented as
\begin{align}
\mathscr L_{g}^{E_g}:
\frac{x + \frac{d^{\text{B},g}}{2}}
{\mathbf k_{g}^{E_g}[1]}
=
\frac{y}
{\mathbf k_{g}^{E_g}[2]}
=
\frac{z}
{\mathbf k_{g}^{E_g}[3]}.
\end{align}

Since the line $\mathscr L_{g}^{E_g}$ passes through a point $\big(-\frac{d^{\text{B},g}}{2}, 0, 0\big)$ inside the ellipsoid $\mathscr E_{g}^{E_g}$, $\mathscr L_{g}^{E_g}$ and ellipsoid $\mathscr E_{g}^{E_g}$ must have two intersections.
Define the coordinate of the intersection point under $\mathcal C^{E_g}$ as $\mathbf p_{\text{S}}^{E_g} = \left(x_0, y_0,z_0\right)$, we have the following equations:
\begin{align}\label{equations_for_intersection}
\left\{
\begin{aligned}
&\frac{x_0^2}{a_{g}^2} + \frac{y_0^2}{b_{g}^2} + \frac{z_0^2}{b_{g}^2} = 1\\
%%%%%%%%%%%%%%%%%%%%%%%%%%%%%%%%%%%%%%%%%%%%%
&\frac{x_0 + \frac{d^{\text{B},g}}{2}}
{\mathbf k_{g}^{E_g}[1]}
= \frac{y_0}
{\mathbf k_{g}^{E_g}[2]}
%\\
%%%%%%%%%%%%%%%%%%%%%%%%%%%%%%%%%%%%%%%%%%%%%%
%&\frac{x_0 + \frac{d^{\text{B},j}}{2}}
%{\mathbf k_{RSU_j}^{E_j}[1]}
= \frac{z_0}
{\mathbf k_{g}^{E_g}[3]}
\end{aligned}
\right.
\end{align}
Thus, the two solutions of $\mathbf p_{\text{S}}^{E_g}$ are $\mathbf p_{\text{S},1}^{E_g}$ and $\mathbf p_{\text{S},2}^{E_g}$ derived respectively by
\eqref{xp_solution_all_1} and \eqref{xp_solution_all_2}
in Appendix A.

In addition, the location of the STAR-RIS should be at the negative direction of $\mathbf k_{g}^{E_g}$ from the left focal point $\mathbf c_1^{E_g} = \big(-\frac{d^{\text{B},g}}{2}, 0, 0\big)$.
Thus we have
%$\mathbf p_{RIS}^{E_j} \in \{\mathbf p_{I,1}^{E_j}, \mathbf p_{I,2}^{E_j}\}$ and $\frac{\left(\mathbf p_{RIS}^{E_j} - \mathbf c_1^{E_j}\right)^T \mathbf k_{RSU_j}^{E_j}}
%{|\mathbf p_{RIS}^{E_j} - \mathbf c_1^{E_j}| |\mathbf k_{RIS}^{E_j}|} = -1$.
\begin{align}
\left\{
\begin{aligned}
&\mathbf p_{\text{S}}^{E_g} \in \{\mathbf p_{S,1}^{E_g}, \mathbf p_{S,2}^{E_g}\} \\
%%%%%%%%%%%%%%%%%%%%%%%%%%%%%%%%%%%%%%%%%%%%%
&\frac{\left(\mathbf p_{\text{S}}^{E_g} - \mathbf c_1^{E_g}\right)^T \mathbf k_{g}^{E_g}}
{\|\mathbf p_{\text{S}}^{E_g} - \mathbf c_1^{E_g}\| \|\mathbf k_{g}^{E_g}\|} = -1
\end{aligned}
\right.
\end{align}
Furthermore, the distance between the STAR-RIS and the $g$-th RSU can be calculated as
$d^{\text{SR}_g} = \|\mathbf p_{\text{S}}^{E_g} - \mathbf p_{g}^{E_g}\|$, where $\mathbf p_{g}^{E_g} = \left(-\frac{d^{\text{B},g}}{2}, 0, 0\right)$.
Then the location of the STAR-RIS under the coordinate system $\mathcal C^{R_g}$ is $\mathbf p_{\text{S}}^{R_g} = d^{\text{SR}_g} \mathbf k_{g}$.
To transform $\mathbf p_{\text{S}}^{R_g}$ into the global coordinate system $\mathcal C^B$, $\mathbf k_{g}$ should be further rotated by
$
\mathbf k_{g}^{B} =  \mathbf R_x^{B} \mathbf R_y^{B} \mathbf R_z^{B} \mathbf k_{g} /
\|\mathbf R_x^{B} \mathbf R_y^{B} \mathbf R_z^{B} \mathbf k_{g}\|
$,
%\begin{align}
%\mathbf k_{g}^{B} =  \frac{\mathbf R_x^{B} \mathbf R_y^{B} \mathbf R_z^{B} \mathbf k_{g}}
%{\|\mathbf R_y^{B} \mathbf R_y^{B} \mathbf R_z^{B} \mathbf k_{g}\|},
%\end{align}
where $\mathbf R_x^{B}$, $\mathbf R_y^{B}$, $\mathbf R_z^{B}$ can be defined similar to
\eqref{rotation_Ejx}-\eqref{rotation_Ejz}.
%\eqref{rotation_x_Ej}-\eqref{rotation_z_Ej}
Note that the rotation angles in $\mathbf R_x^{B}$, $\mathbf R_y^{B}$, $\mathbf R_z^{B}$ can be previously set when deploying the RSUs.
Hence, the location of the STAR-RIS under $\mathcal C^B$ is $\widehat{\mathbf p}_{\text{S}}^{g} = \mathbf p_{g} - d^{\text{SR}_g}\mathbf k_{g}^{B}$.
%Besides, since the AODs of the LOS path $\{\widehat{\theta}_{0}^{\text{B}}, \widehat{\phi}_{0}^{\text{B}}\}$ from the BS are also obtained at each RSU, it can also be utilized for the localization in the same way.
%Then we can obtain the result as $\widehat{\mathbf p}_{RIS}^{g,\text{B}}$
%Thus, the localization result at the $g$-th RSU can be refined as $\widehat{\mathbf p}_{RIS}^{g} = (\widehat{\mathbf p}_{RIS}^{g,g} + \widehat{\mathbf p}_{RIS}^{g,\text{B}})/2$.
Furthermore, averaging the localization results of all the RSUs, the STAR-RIS's final location can be derived as $\widehat{\mathbf p}_{\text{S}} = \frac{1}{G} \sum\limits_{g=1}^{G} \widehat{\mathbf p}_{\text{S}}^{g}$.

%\textcolor[rgb]{1.00,0.00,0.00}{Besides, to refine the localization result, we can remove one estimate $\widehat{\mathbf p}_{RIS}^{\text{B},j'}$ which has the largest distance with $\widehat{\mathbf p}_{RIS}^{\text{B}}$, and update $\widehat{\mathbf p}_{RIS}^{\text{B}}$ as $\widehat{\mathbf p}_{RIS}^{\text{B}} = \frac{1}{J} \sum_{j \ne j'} \widehat{\mathbf p}_{RIS}^{\text{B},j}$.}

\begin{remark}
Note that we are considering a very general case that $\mathbf k_{g}^{E_g}[1]$, $\mathbf k_{g}^{E_g}[2]$, and $\mathbf k_{g}^{E_g}[3] $ are all not equal to zero.
If any element of $\mathbf k_{g}^{E_g}$ is equal to zero, $\mathscr L_{g}^{E_g}$ will be perpendicular to one or two axes of $\mathcal C^{E_g}$.
Hence, some elements of $\mathbf p_{S}^{E_g}$ will be determined directly, and \eqref{equations_for_intersection} will degenerate to a problem with less parameters, which will become easier to be solved.
\end{remark}

\begin{remark}
After the vehicle localization, the BS precoder and the RSUs combiner can be designed towards the direction of the LOS path with respect to the STAR-RIS.
For parameter extraction in later frames, the range of beam scanning can be shrunk.
Thus, the number of training sequences can be extremely reduced, which can significantly shorten the training overhead.
\end{remark}

\subsection{Vehicle Velocity Measurement}

From parameter extraction, we obtained the cascaded Doppler frequency shift of the LOS path for BS-RIS-RSUs links, denoted by $\widehat{\nu}_{1}^{\text{B},g}$.
Besides, define the vehicle velocity as $\mathbf v_V \in \mathbb R^{3 \times 1}$ under $\mathcal C^B$, we have
\begin{align}\label{nu_rep}
\widehat{\nu}_{1}^{\text{B},g} = \frac{1}{\lambda}\|\mathbf v_V\|\cos{\gamma_B} + \frac{1}{\lambda}\|\mathbf v_V\|\cos{\gamma_g},
\end{align}
where ${\gamma_B}$ is the angle between the direction of $\mathbf v_V$ and the AOA at the STAR-RIS from the BS, and ${\gamma_g}$ is the angle between the direction of $\mathbf v_V$ and the AOD from the STAR-RIS to the $g$-th RSU.
Since we have obtained the location of the vehicle, $\cos{\gamma_B}$ and $\cos{\gamma_g}$ can be derived as
\begin{align}
\cos{\gamma_B} \!=\! \frac{(\widehat{\mathbf p}_{\text{S}}-\!\mathbf p_{\text{B}})^T \mathbf v_V}{\|\widehat{\mathbf p}_{\text{S}}-\!\mathbf p_{\text{B}}\| \!\cdot\! \|\mathbf v_V\|},
\cos{\gamma_g} \!=\! \frac{(\widehat{\mathbf p}_{\text{S}}-\!\mathbf p_{g})^T \mathbf v_V}{\|\widehat{\mathbf p}_{\text{S}}-\!\mathbf p_{g}\| \!\cdot\! \|\mathbf v_V\|}.
\end{align}
respectively.
Thus, \eqref{nu_rep} can be rewritten as
\begin{align}\label{nu_rep2}
\widehat{\nu}_{1}^{\text{B},g}
= \Big(\frac{\widehat{\mathbf p}_{\text{S}}-\mathbf p_{\text{B}}}{\|\widehat{\mathbf p}_{\text{S}}-\mathbf p_{\text{B}}\| \lambda}
+ \frac{\widehat{\mathbf p}_{\text{S}}-\mathbf p_{g}}{\|\widehat{\mathbf p}_{\text{S}}-\mathbf p_{g}\| \lambda}\Big)^T
\mathbf v_V.
\end{align}
Define $\mathbf q_g  = \frac{\widehat{\mathbf p}_{\text{S}}-\mathbf p_{\text{B}}}{\|\widehat{\mathbf p}_{\text{S}}-\mathbf p_{\text{B}}\| \lambda}
+ \frac{\widehat{\mathbf p}_{\text{S}}-\mathbf p_{g}}{\|\widehat{\mathbf p}_{\text{S}}-\mathbf p_{g}\| \lambda}$, $\mathbf Q \!=\! [\mathbf q_1, \mathbf q_2, \ldots, \mathbf q_G]$,
and
$\widehat{\boldsymbol \nu}_{1} \!=\! [\widehat{\nu}_{1}^{\text{B},1}, \widehat{\nu}_{1}^{\text{B},2}, \ldots, \widehat{\nu}_{1}^{\text{B},G}]^T$.
By stacking the equations about all the Doppler frequency shifts $\{\widehat{\nu}_{1}^{\text{B},g} \}_{g=1}^{G}$, we have
$
\mathbf Q^T \mathbf v_V = \widehat{\boldsymbol \nu}_{1}
$,
%\begin{align}\label{nu_solve}
%\mathbf Q^T \mathbf v_V = \widehat{\boldsymbol \nu}_{0},
%\end{align}
and $\mathbf v_V$ can be finally obtained as
$
\widehat{\mathbf v}_V = \mathbf Q \left(\mathbf Q^{T} \mathbf Q\right)^{-1}
\widehat{\boldsymbol \nu}_{1}
$.
%\begin{align}\label{nu_ans}
%\widehat{\mathbf v}_V = \mathbf Q \left(\mathbf Q^{T} \mathbf Q\right)^{-1}
%\widehat{\boldsymbol \nu}_{0}.
%\end{align}
With the acquired user velocity and location, we can further predict the user location in the later transmission frames.

\subsection{STAR-RIS Reflection and Refraction Design and Prediction}

The performance of sensing is highly related to the accuracy of the estimated parameters at each RSU.
Thus, the STAR-RIS reflection phase shift matrix, also known as the phase shift matrix $\mathbf \Omega_{O}$ of the outside omni-RIS with a certain reflection energy splitting factor $\epsilon_R^O$, should be designed to enhance the received signal strength at the RSUs.
Since the training sequences for each RSU are the same, we can separately design the phase shift matrix with respect to each RSU, and then sum up all $G$ designed matrices to build the overall STAR-RIS reflection phase shift matrix.

Assume that the orientation of the STAR-RIS is determined by the velocity direction of the vehicle.
Since the velocity and the location of the vehicle has been sensed, we can acquire the rotation matrices from the global coordinate system to the STAR-RIS local coordinate system, denoted by $\mathbf R_x^S$, $\mathbf R_y^S$, and $\mathbf R_z^S$.
Then, the direction vector from the vehicle to the RSUs can be rotated into $\mathcal C^S$ by
\begin{align}
\mathbf k_{g}^{S} = \frac{\mathbf R_x^{S} \mathbf R_y^{S} \mathbf R_z^{S} (\mathbf p_{g} - \widehat{\mathbf p}_{\text{S}})}
{\|\mathbf R_x^{S} \mathbf R_y^{S} \mathbf R_z^{S} (\mathbf p_{g} - \widehat{\mathbf p}_{\text{S}})\|}.
\end{align}
Then, the azimuth and elevation AODs from the STAR-RIS to the RSUs can be calculated as
\begin{align}
\widehat{\theta}^{\text{SR}_g} =&
\arccos \frac{\Big(\mathbf k_{g}^{S} - \frac{(\mathbf k_{g}^{S})^T \mathbf e_{z}^{S}}{\|\mathbf e_{z}^{S}\|} \mathbf e_{z}^{S}\Big)^T \mathbf e_{x}^{S}}
{\big\|\mathbf k_{g}^{S} - \frac{(\mathbf k_{g}^{S})^T \mathbf e_{z}^{S}}{\|\mathbf e_{z}^{S}\|} \mathbf e_{z}^{S}\big\| \cdot \left\|\mathbf e_{x}^{S}\right\|},
\\
\widehat{\phi}^{\text{SR}_g} =&
\frac{\pi}{2} - \arccos\frac{(\mathbf k_{g}^{S})^T \mathbf e_{z}^{S}}{\|\mathbf k_{g}^{S}\| \cdot \|\mathbf e_{z}^{S}\|}.
\end{align}
where $\mathbf e_{x}^{S}$ and $\mathbf e_{z}^{S}$ are the direction vector along the $x$ and $z$ axes of $\mathcal C^S$, respectively.
The AOAs $\{\widehat{\theta}^{\text{BS}}_{1}, \widehat{\phi}^{\text{BS}}_{1}\}$ of the LOS path at the STAR-RIS from the BS can be calculated in the same way.
However, the AOAs of NLOS paths at the STAR-RIS can not be decoupled in this circumstance.
Thus, we turn to design the phase shift matrix of the STAR-RIS for each RSU with only the LOS cascaded path.
With the parameters related to the STAR-RIS reflection for the $g$-th RSU, the problem for the sub-optimal reflection phase shift matrix can be derived by
\begin{align}
\overline{\mathbf \Omega}_{O}^g = \arg\max\limits_{\mathbf \Omega_O^g} \left|\mathbf{a}_{\text{S}}^T (\widehat{\theta}^{\text{SR}_g}, \widehat{\phi}^{\text{SR}_g})
\mathbf \Omega_O^g
\mathbf{a}_{\text{S}} (\widehat{\theta}^{\text{BS}}_{1}, \widehat{\phi}^{\text{BS}}_{1})\right|.
\end{align}
Then the phase shift vector $\boldsymbol \omega_O$ can be solved and normalized as
\begin{align}
[\overline{\boldsymbol \omega}_O^g]_{n_s}
=
\frac{[\mathbf{a}_{\text{S}} (\widehat{\theta}^{\text{SR}_g}, \widehat{\phi}^{\text{SR}_g})]_{n_s}^*
[\mathbf{a}_{\text{S}} (\widehat{\theta}^{\text{BS}}_{1}, \widehat{\phi}^{\text{BS}}_{1})]_{n_s}^* }
{\left|[\mathbf{a}_{\text{S}}^* (\widehat{\theta}^{\text{SR}_g}, \widehat{\phi}^{\text{SR}_g})]_{n_s}
[\mathbf{a}_{\text{S}}^* (\widehat{\theta}^{\text{BS}}_{1}, \widehat{\phi}^{\text{BS}}_{1})]_{n_s} \right|},
\end{align}
where $n_s = 0,1,\ldots, N_S-1$.
By summing up and normalizing the designed phase vector for all the RSUs, the overall reflection phase shift vector can be obtained as
$
\widetilde{\boldsymbol \omega}_O
=
\sum\limits_{g=1}^{G}
\overline{\boldsymbol \omega}_O^g
/
\|\sum\limits_{g=1}^{G}
\overline{\boldsymbol \omega}_O^g\|
$.
%\begin{align}\label{reflection_design_final}
%\widetilde{\boldsymbol \omega}_O
%=
%\frac{\sum\limits_{g=1}^{G}
%\overline{\boldsymbol \omega}_O^g}
%{\left\|\sum\limits_{g=1}^{G}
%\overline{\boldsymbol \omega}_O^g\right\|}.
%\end{align}

For the STAR-RIS refraction design,
since the space in the vehicle is very small, we assume that the UE's location with respect to the STAR-RIS is known by searching in advance \cite{nearfield_dai}.
Besides, since the phase shift matrix of the outside RIS has been determined, the refraction design is equivalent to design the phase shift matrix $\mathbf \Omega_I$ of the inside omni-RIS.
Then, the sub-optimal phase shift of the inside omni-RIS can be similarly derived by
\begin{align}\label{in_RIS_design_problem}
\widetilde{\boldsymbol \omega}_{I} = \arg\max\limits_{{\boldsymbol \omega}_{I}} \left|\mathbf{a}_{\text{SU}}^T
\mathbf \Omega_I
\mathbf G
\widetilde{\mathbf \Omega}_O
\mathbf{a}_{\text{S}} (\widehat{\theta}^{\text{BS}}_{1}, \widehat{\phi}^{\text{BS}}_{1})\right|,
\end{align}
which can be solved as
\begin{align}\label{refraction_design_final}
[\widetilde{\boldsymbol \omega}_I]_{n_s}
=
\frac{[\mathbf{a}_{\text{SU}}]_{n_s}^*
[\mathbf G
\widetilde{\mathbf \Omega}_O
\mathbf{a}_{\text{S}} (\widehat{\theta}^{\text{BS}}_{1}, \widehat{\phi}^{\text{BS}}_{1})]_{n_s}^* }
{\left|[\mathbf{a}_{\text{SU}}]_{n_s}
[\mathbf G
\widetilde{\mathbf \Omega}_O
\mathbf{a}_{\text{S}} (\widehat{\theta}^{\text{BS}}_{1}, \widehat{\phi}^{\text{BS}}_{1})]_{n_s} \right|},
\end{align}
where $n_s = 0,1,\ldots, N_S-1$.
Moreover, with the acquired vehicle location and velocity, we can predict the vehicle's location at the next sensing duration.
Thus, we can pre-design the reflection and refraction phase shift vectors in the same way to enhance the received signal strength at the RSUs and the UE.
\begin{remark}
If we simultaneously scan the BS precoder, the RSU combiner, and the STAR-RIS reflecting phase shift matrix in the training process, the AOAs and AODs at the STAR-RIS can be decoupled and be utilized for the optimal STAR-RIS reflection design without sensing results.
However, since RIS always consists of enormous elements, the codebook of the STAR-RIS phase shift will be very large.
This will incur huge training overhead.
Thus, we turn to the sub-optimal STAR-RIS reflection phase shift matrix design for reducing the cost of parameter extraction.
Similar reason and strategy are considered for the refraction phase shift matrix design.

\end{remark}

\subsection{Trade-off Design for Sensing and Communication}
Even though the phase shift vectors of the reflection and refraction have been designed, we should optimize the reflection and refraction energy splitting factors, i.e., $\epsilon_R^O$ and $\epsilon_T^O$, so that we can achieve a balanced trade-off between the sensing and communication performance.
Since the sensing accuracy and communication quality are highly related to the received signal-to-noise ratio (SNR), we aim to reach a trade-off with respect to the SNRs at the RSUs and the UE.

From the received signal model at the RSUs and the UE, the SNR of STAR-RIS related received signal can be derived as
%\begin{align}
%\text{SNR}_g
%%=&
%%\frac{\mathbb E\left\{\sum\limits_{p=1}^{P} \left\| h_p^{\text{B},g} \mathbf{a}_{\text{RSU}} (\theta^{g}, \phi^{g})
%%\mathbf{a}_{\text{R}}^T (\theta^{\text{SR}_g}, \phi^{\text{SR}_g})
%%\widetilde{\mathbf \Omega}_R
%%\mathbf{a}_{\text{R}} (\theta^{\text{BS}}_{p}, \phi^{\text{BS}}_{p})
%%\right\|^2\right\}\sigma_P^2
%%}
%%{\sigma_N^2}
%%\notag\\
%=&
%\frac{1}{\sigma_N^2}
%\Big|\sum\limits_{p=1}^{P}
%\mathbf{a}_{\text{R}}^T (\theta^{\text{SR}_g}, \phi^{\text{SR}_g})
%\widetilde{\mathbf \Omega}_O
%\mathbf{a}_{\text{R}} (\theta^{\text{BS}}_{p}, \phi^{\text{BS}}_{p})
%\Big|^2 \epsilon_R^O \lambda_p^{\text{B},g} \sigma_P^2
%,
%\end{align}
%and
%\begin{align}
%\text{SNR}_{\text{UE}}
%=&
%\frac{1}{\sigma_N^2}
%\Big|\sum\limits_{p=1}^{P}
%\mathbf{a}_{\text{US}}^T
%\widetilde{\mathbf \Omega}_I \mathbf G \widetilde{\mathbf \Omega}_O
%\mathbf{a}_{\text{R}} (\theta^{\text{BS}}_{p}, \phi^{\text{BS}}_{p})
%\Big|^2 \epsilon_{T}^O \epsilon_{T}^I \lambda_p^{\text{B},g} \sigma_P^2
%.
%\end{align}
\begin{align}
\text{SNR}_g\!\!
=&\!
\frac{1}{\sigma_N^2}\!
\Big|\!\!\sum\limits_{p=1}^{P}\!\!
\mathbf{a}_{\text{S}}^T \!(\theta^{\text{SR}_g}\!\!, \phi^{\text{SR}_g}\!)
\widetilde{\mathbf \Omega}_O
\mathbf{a}_{\text{S}} \!(\theta^{\text{BS}}_{p}\!, \phi^{\text{BS}}_{p}\!)
\Big|^2
\!\!\!\epsilon_R^O \lambda_p^{\text{B},g} \!\sigma_P^2
, \\
\text{SNR}_{\text{UE}}\!\!
=&\!
\frac{1}{\sigma_N^2}\!
\Big|\!\!\sum\limits_{p=1}^{P}\!\!
\mathbf{a}_{\text{SU}}^T
\widetilde{\mathbf \Omega}_I \mathbf G \widetilde{\mathbf \Omega}_O
\mathbf{a}_{\text{S}} (\theta^{\text{BS}}_{p}, \phi^{\text{BS}}_{p})
\Big|^2
\!\!\epsilon_{T}^O \epsilon_{T}^I \lambda_p^{\text{B},g} \sigma_P^2
\end{align}
respectively.
Since the phase shift matrices are developed for the LOS path, the power of the NLOS paths will be very small and can be neglected due to the beam focusing characteristic of RIS.
Hence, the SNRs can be approximated as
\begin{align}
\text{SNR}_g\!\!
\approx&
\frac{1}{\sigma_N^2}\!\!
\left|
\mathbf{a}_{\text{S}}^T \!(\theta^{\text{SR}_g}, \phi^{\text{SR}_g}\!)
\widetilde{\mathbf \Omega}_O
\mathbf{a}_{\text{S}} (\theta^{\text{BS}}_{1}, \phi^{\text{BS}}_{1})
\!\right|^2
\!\!\!\epsilon_R^O \lambda_{1}^{\text{B},g} \sigma_P^2
,
\\%%%%%%%%%%%%%%%%%%%%%%%%%%%%%%%%%%%%%%%%%%%%%%%%%%%%%%%%%%%%%%
\text{SNR}_{\text{UE}}\!\!
\approx&
\frac{1}{\sigma_N^2}\!\!
\left|
\mathbf{a}_{\text{SU}}^T
\widetilde{\mathbf \Omega}_I \mathbf G \widetilde{\mathbf \Omega}_O
\mathbf{a}_{\text{S}} (\theta^{\text{BS}}_{1}, \phi^{\text{BS}}_{1})
\!\right|^2
\!\!\!\epsilon_{T}^O \epsilon_{T}^I \lambda_{1}^{\text{BS}} \sigma_P^2
.
\end{align}

Without loss of generality, the sensing performance is determined by the RSU with the smallest SNR, denoted by $\text{SNR}_{g^{\text{min}}}$.
Thus, the trade-off design can be reached by
\begin{align}\label{opt_factor1}
\arg\max\limits_{\epsilon_R^O, \epsilon_T^O} \ \min  \left\{\kappa_S \text{SNR}_{g^{\text{min}}}
, \kappa_C \text{SNR}_{\text{UE}} \right\},
\end{align}
where $\kappa_S$ and $\kappa_C$ are the weight factors of SNRs for sensing and communication, respectively, and $\kappa_S + \kappa_C = 1$.
Then, \eqref{opt_factor1} can be transformed to
\begin{align}
\arg\min\limits_{\epsilon_R^O, \epsilon_T^O}
\left|\kappa_S \text{SNR}_{g^{\text{min}}}
-\kappa_C \text{SNR}_{\text{UE}}
\right|.
\end{align}
Hence, we can obtain $\epsilon_R^O$ as \eqref{solution_epsilon_RO} at the top of the next page
\begin{figure*}[htbp]
\begin{align}\label{solution_epsilon_RO}
\epsilon_R^O =
\frac{\kappa_C
\left|
\mathbf{a}_{\text{SU}}^T
\widetilde{\mathbf \Omega}_I \mathbf G \widetilde{\mathbf \Omega}_O
\mathbf{a}_{\text{S}} (\theta^{\text{BS}}_{1}, \phi^{\text{BS}}_{1})
\right|^2 \epsilon_{T}^I \lambda_{1}^{\text{BS}}}
{\kappa_S \left|
\mathbf{a}_{\text{S}}^T (\theta^{\text{SR}_{g^{\text{min}}}}, \phi^{\text{SR}_{g^{\text{min}}}})
\widetilde{\mathbf \Omega}_O
\mathbf{a}_{\text{S}} (\theta^{\text{BS}}_{1}, \phi^{\text{BS}}_{1})
\right|^2 \lambda_{1}^{\text{B},{g^{\text{min}}}}
\!\!\!+\!\!
\kappa_C \!
\left|
\mathbf{a}_{\text{SU}}^T
\widetilde{\mathbf \Omega}_I \mathbf G \widetilde{\mathbf \Omega}_O
\mathbf{a}_{\text{S}} (\theta^{\text{BS}}_{1}, \phi^{\text{BS}}_{1})
\right|^2
\!\epsilon_{T}^I \lambda_{1}^{\text{BS}}},
\end{align}
\hrulefill
\end{figure*}
and $\epsilon_T^O = 1 - \epsilon_R^O$.

\section{Simulation Results}

In this section, we evaluate the performance of our proposed ISAC scheme through numerical simulation.
In our ISAC system, we consider an area of $400 \text{m}\times 400 \text{m} \times 8 \text{m}$.
The height of BS, RSUs, and the vehicle are set as $8$ m, $4$ m, and $1.5$ m, respectively.
The BS, the STAR-RIS, and the RSUs are all equipped with $16 \times 16$ UPA.
Besides, the number of RF chains $N_B^{RF} $ at the BS and $N_R^{RF}$ at the RSUs are both set as $32$.
%Thus, \textcolor[rgb]{1.00,0.00,0.00}{$L_O = \frac{16 \times 16}{32} \times \frac{16 \times 16}{32} = 64$ pilot sequences} are needed to scan all the beams within a DFT-based codebook.
The carrier frequency is 30 GHz, and the antenna spacing $d$ is set as half of the wavelength.
The vehicle moves at speed $|\mathbf v_V| = 80$ km/h, and thus the maximal Doppler frequency shift is $2.222$ kHz.
The system sampling rate is $T_s = \frac{1}{100 \,  \text{MHz}}$, and the length of the pilot sequence is $N_T = 1200$.
The number of scattering paths from the BS to the vehicle is $P=3$, while that from the BS to the RSUs are $P_g = P + 3, g = 1,2,\ldots, G$.
Moreover, the exponentially decaying power delay profile $\sigma_{h_p}^2 = \sigma_c^2 e^{-\frac{\tau_p}{\tau_{max}}}$ is adopted for $h^{\text{BS}}_p$ and $h^{\text{B},g}_p$, respectively, where the constant $\sigma_c^2$ is chosen such that the average cascaded channel power is normalized to unity.
The maximum delay spread is set as $\tau_{max} = 10 T_s$.
Here, we use the normalized mean square error (NMSE) for the estimated parameters, which is defined as
$
\text{NMSE}_{\mathbf x} = \mathbb E
\left\{ \frac{\|\widehat{\mathbf x} - \mathbf x\|^2}{\|\mathbf x\|^2}
 \right\}, \mathbf x = \boldsymbol \nu$ or $\overline{\mathbf h}$
with $\widehat{\mathbf x}$ representing the estimation of $\mathbf x$.
The sensing performance is evaluated by the root-mean-square-error (RMSE), which is defined as $
\text{RMSE}_{\mathbf x} \!=\! \sqrt{\mathbb E
\left\{ \|\widehat{\mathbf x} - \mathbf x\|^2
 \right\}}$,
where $\mathbf x = \mathbf p_{\text{S}}$ or $\mathbf v_V$.

\subsection{Performance of Parameter Extraction}

Firstly, we examine the performance of MOMP based parameter extraction.
Since the AOAs and AODs can be directly obtained by choosing the scanning beams of the received sequences with highest power, we only test the estimation accuracy of Doppler frequency shift and the equivalent channel gain.

\begin{figure}[htbp]
 \centering
 \includegraphics[width=80mm]{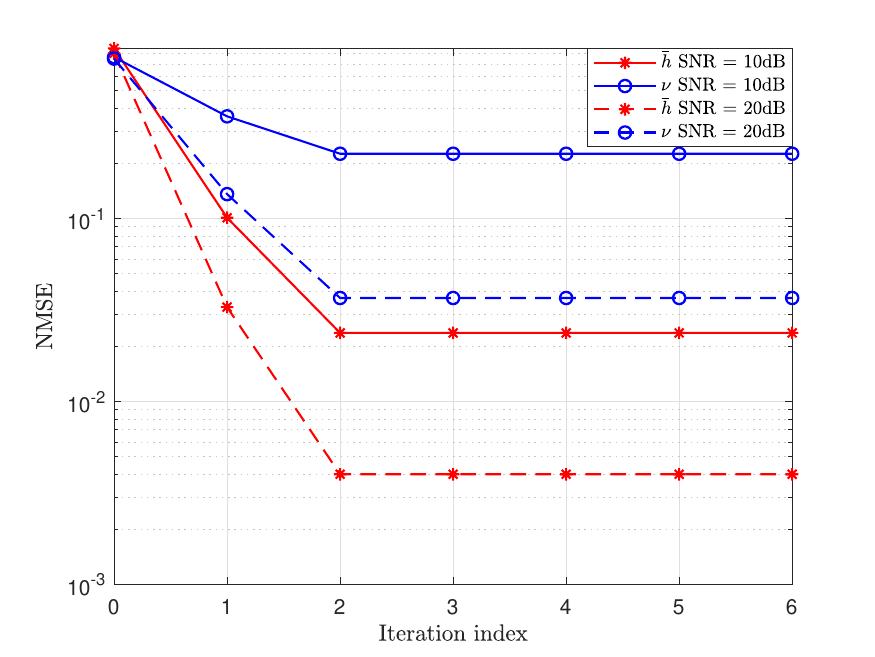}
 \caption{NMSEs versus MOMP iterations.}
 \label{MSE_vs_iter_2SNR}
\end{figure}

\figurename{ \ref{MSE_vs_iter_2SNR}} shows the NMSE of the channel parameters versus the MOMP iteration index, where $\overline{h}$ is the equivalent channel gain, and $\nu$ is the cascaded Doppler frequency shift.
Note that the iteration index ``0'' means that the searched initial points are directly regarded as the final searching results.
It can be observed that the NMSEs decrease with the increase of itertion times, and approach their steady states after $2$ iterations.
This result efficiently verifies the convergence and effectiveness of the proposed MOMP based parameter extraction.

\begin{figure}[htbp]
 \centering
 \includegraphics[width=80mm]{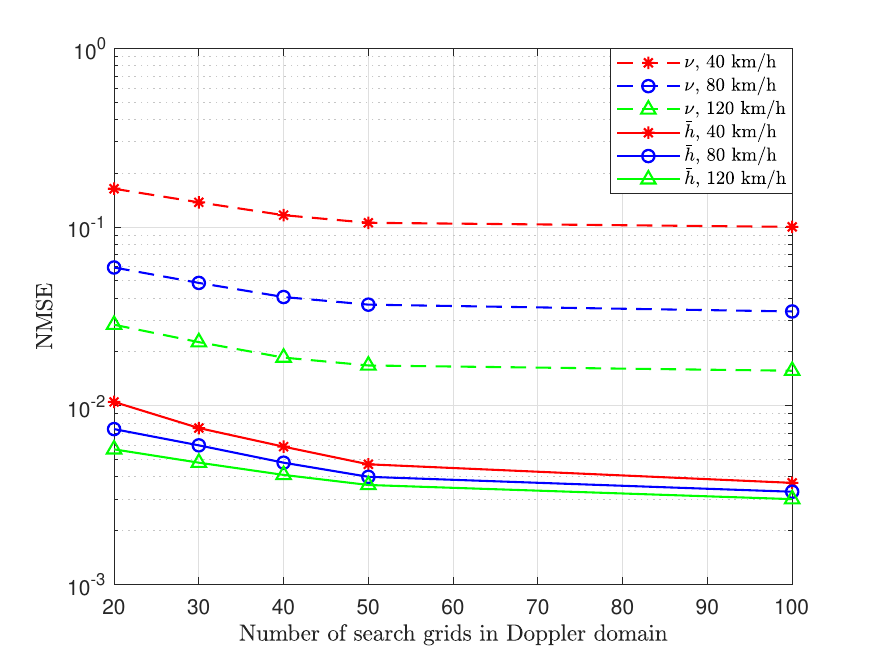}
 \caption{NMSEs versus the number of Doppler searching grids.}
 \label{MSE_vs_Dopnum}
\end{figure}

In \figurename{ \ref{MSE_vs_Dopnum}}, the NMSEs versus the number of search grids in Doppler domain are presented, where the SNR is set as $20$ dB, and 3 different velocities are considered.
With the increase of search grids, the NMSEs decrease first and converge after a certain number of search grids in Doppler domain.
Thus, enlarging the number of search grids can not noticeably improve the performance.
Furthermore, if coarse estimation of the parameters are obtained by previous training processes, the searching range for each scattering path can be shrunk to reduce the searching complexity.
In the rest of our experiments, we shall use 50 search grids in Doppler domain.
Moreover, the estimation performance improves as the velocity increases.
This is due to the fact that higher velocity results in larger Doppler frequency shift, which can be more easier to be observed.

\begin{figure}[htbp]
 \centering
 \includegraphics[width=80mm]{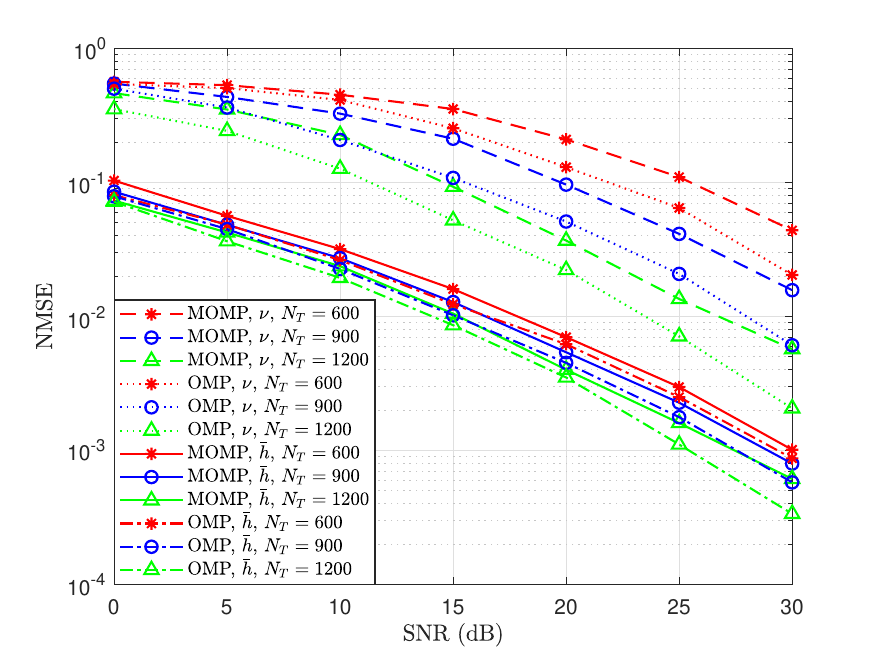}
 \caption{NMSEs versus SNR.}
 \label{MSE_vs_SNR}
\end{figure}

\figurename{ \ref{MSE_vs_SNR}} illustrates the NMSEs versus SNR for different lengths of training sequences under velocity of $80$ km/h, where $50$ search grids in Doppler domain are considered.
From \figurename{ \ref{MSE_vs_SNR}}, we can observe that the longer the training sequence, the lower the NMSEs of the estimation.
This can be explained as follows.
With longer training sequence, the phase cumulation caused by Doppler frequency shift will be larger, which can be more easier to be recognized and estimated.
Besides, we can also observe that the NMSEs decrease with the increase of SNR.
It can be further noticed that the NMSEs with lower SNR decrease slower than that with higher SNR.
This is because that the phase shifts caused by Doppler frequency shift are not very noticeable.
Thus, the searching accuracy of Doppler frequency shift is dominated by the strength of noise.
Moreover, \figurename{ \ref{MSE_vs_SNR}} also shows the comparison between the performance of OMP algorithm \cite{OMP_theory} and the utilized MOMP algorithm.
It can be checked that the NMSEs of the parameters with MOMP are slightly higher than that with OMP under different length of training sequence.

However, due to the joint search of multiple parameters, OMP requires a huge searching map, and results in higher computational complexity.
On the other hand, in MOMP, the
parameters of a scattering path are searched one by one, which can reduce the computational complexity significantly \cite{MOMP}.
Specific complexity comparison for OMP and MOMP except for beam searching is shown in \tablename{ \ref{table_comp_OMP_MOMP}}, where $\eta_{\nu}$ and $\eta_{\tau}$ are the oversampling rates respectively for Doppler frequency shift and delay.
When the requirement of parameter extraction resolution is higher, $\eta_{\nu}$ and $\eta_{\tau}$ should be larger, and MOMP exhibits lower computational complexity than OMP.
\begin{table}[!t]
	\centering
	\renewcommand{\arraystretch}{1.2}
	\caption{Computational complexity comparison between OMP and MOMP.}
	\label{table_comp_OMP_MOMP}
\begin{tabular}{|c|l|}
  \hline
  Algorithm & Computational complexity \\\hline
  OMP \cite{OMP_theory} & $\mathcal{O}( (N_{\nu}N_{\tau})^2 \eta_{\nu}\eta_{\tau} )$ \\\hline
  MOMP & $\mathcal{O}( (N_{\nu}N_{\tau})^2 + (1+N_{\nu}N_{\tau})(N_{\nu}^2 \eta_{\nu} + N_{\tau}^2 \eta_{\tau}) )$ \\
  \hline
\end{tabular}
\end{table}

\subsection{Performance of Localization and Velocity Measurement}

\begin{figure}[htbp]
 \centering
 \includegraphics[width=80mm]{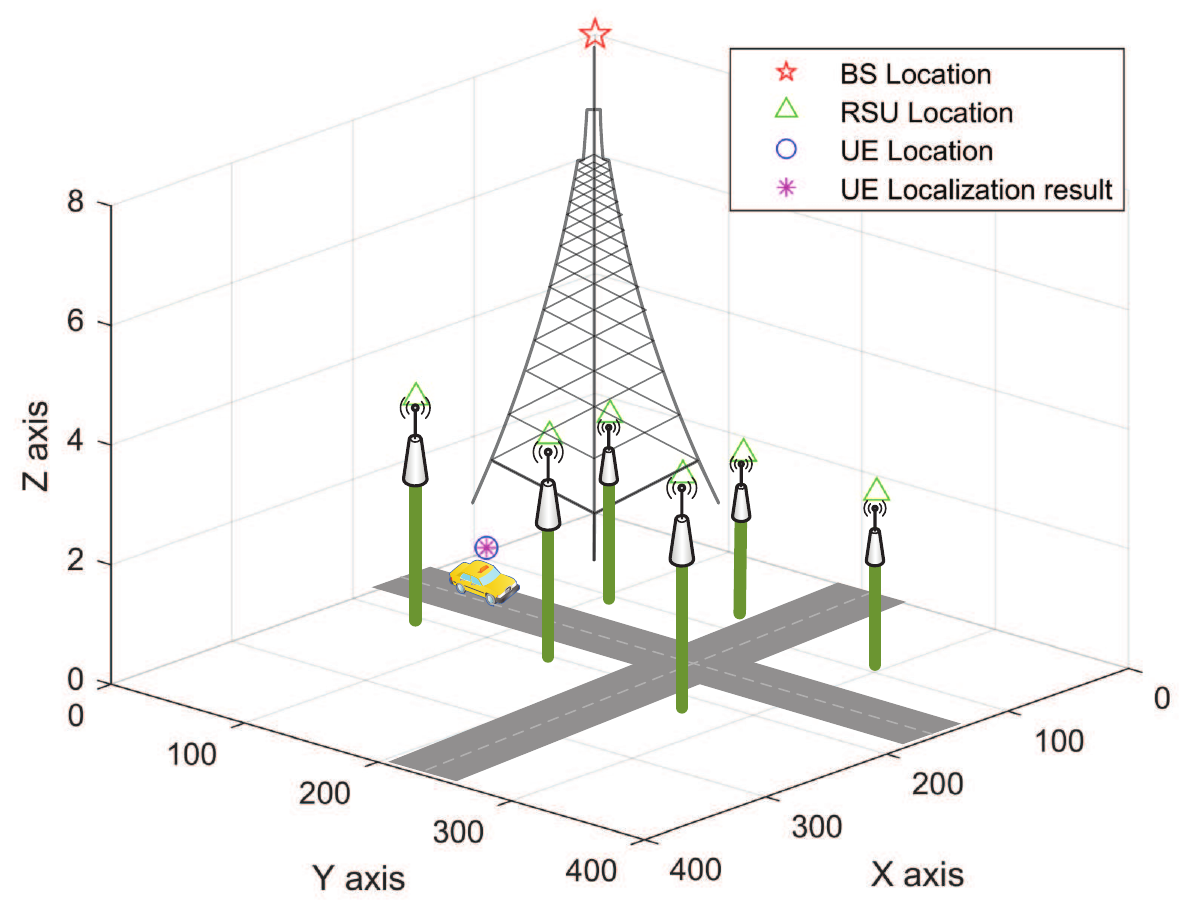}
 \caption{Illustration of vehicle localization.}
 \label{illustration_of_localization}
\end{figure}

Then, we focus on the performance of sensing.
First, we verify the feasibility of our proposed localization scheme with precise channel parameters in \figurename{ \ref{illustration_of_localization}}, where $6$ RSUs are considered.
It can be found that the proposed scheme can work very well with precise channel parameters.

\begin{figure}[htbp]
 \centering
 \includegraphics[width=80mm]{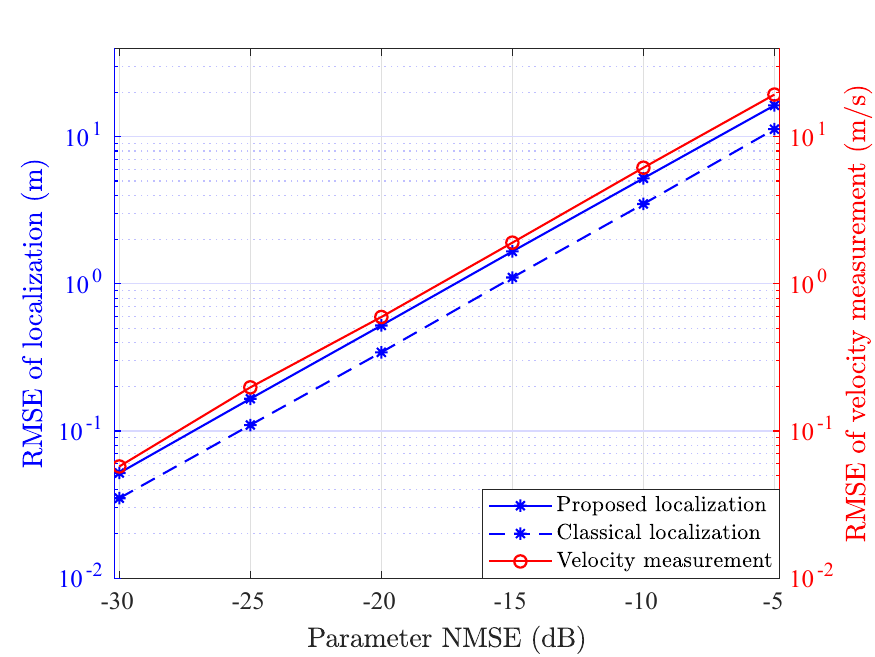}
 \caption{Sensing error versus the NMSE of channel parameters.}
 \label{Sensing_error_vs_para_NMSE}
\end{figure}

%We first focus on the performance of sensing under different circumstances.
Then, we will discuss the performance of localization and velocity measurement under different system settings.
In \figurename{ \ref{Sensing_error_vs_para_NMSE}}, we exhibit the RMSEs of localization and velocity measurement versus the parameters' estimation NMSE, where $6$ RSUs are considered.
Note that the parameter related to localization lies on the angles, while that related to velocity measurement lies on the Doppler frequency shifts.
It can be observed that the RMSEs of both localization and velocity measurement increase almost linearly as the NMSEs of the parameters increase.
%When parameter NMSE is $-15$ dB, it can be observed that the absolute error of proposed localization is only less than $2$ m, and that of velocity measurement is only $2$ m/s, which are very small and acceptable.

\begin{figure}[htbp]
 \centering
 \includegraphics[width=80mm]{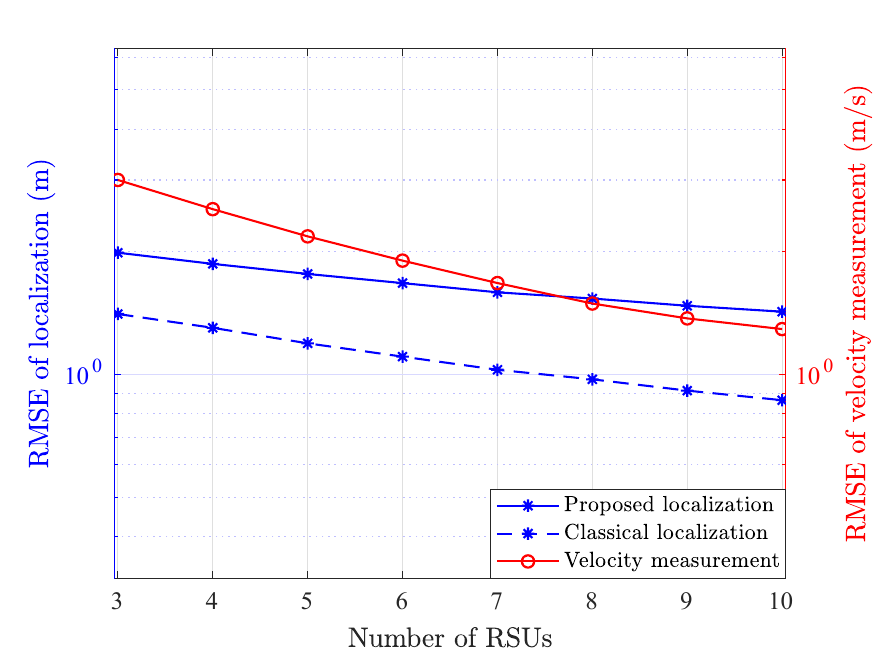}
 \caption{Sensing error versus the number of RSUs.}
 \label{Sensing_error_vs_RSUnum}
\end{figure}

\figurename{ \ref{Sensing_error_vs_RSUnum}} illustrates the RMSEs of localization and velocity measurement versus the number of RSUs, where the channel parameters NMSEs are set as $-15$ dB.
With the number of RSU increases, the RMSEs of both localization and velocity measurement decrease.
Besides, when the number of RSU gets higher, the sensing errors tend to converge at a very low level, which shows the effectiveness of our proposed sensing scheme in high mobility scenario.

We also compare the performance of our proposed geometry aided localization with the classical parameter-only localization \cite{Shi_jin_loc} in \figurename{ \ref{Sensing_error_vs_para_NMSE}} and \figurename{ \ref{Sensing_error_vs_RSUnum}}.
It can be observed that the performance of the proposed method is slightly weaker than that of the classical one.
However, the parameter-only localization requires the decoupling of parameters about the BS-RIS and the RIS-RSU links, leading to additional computational complexity for the estimation of double-sides parameters.
By contrast, since the proposed scheme only utilizes the cascaded parameters, which can simplify the ISAC protocol and reduce the parameter extraction overhead significantly.

\subsection{Trade-off Design for Sensing and Communication}

\begin{figure}[htbp]
 \centering
 \includegraphics[width=80mm]{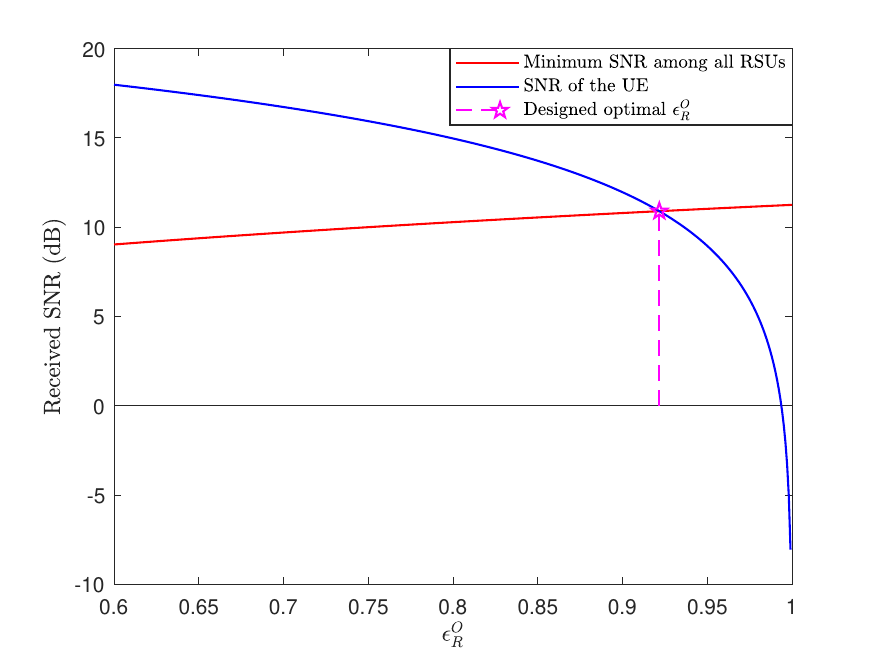}
 \caption{Received SNR at the UE and the worst RSU versus $\epsilon_R^O$.}
 \label{Design_of_epsilon}
\end{figure}

Finally, we verify the proposed trade-off design scheme for sensing and communication.
Note that the phase shift matrices and energy splitting factors of the STAR-RIS are optimized in Sections IV.D and E.
As illustrated in \figurename{ \ref{Design_of_epsilon}}, with the increase of $\epsilon_R^O$, the minimum received SNR of the RSUs increases, and the received SNR of the UE decreases.
It can be found that the optimized $\epsilon_R^O$ in Section IV.E is just on the intersection point, which verifies the effectiveness of the proposed trade-off design scheme.

\section{Conclusion}

In this paper, we proposed a novel STAR-RIS aided ISAC scheme over high mobility scenario, where a STAR-RIS was equipped on the outside surface of a vehicle.
Firstly, we developed an efficient transmission structure for the ISAC scheme, where a number of training sequences with orthogonal precoders and combiners were respectively utilized at BS and RSUs for channel parameter extraction.
Then, we characterized the near-field static channel model between the STAR-RIS and in-vehicle UE as well as the far-field time-frequency selective BS-RIS-RSUs channel model.
By implementing MOMP algorithm, the cascaded channel parameters of the scattering paths for BS-RIS-RSUs links were obtained.
With the joint utilization of these extracted cascaded channel parameters from all the RSUs, the vehicle localization and its velocity measurement were obtained.
Benefiting from the sensing results, the reflection and refraction phase shifts of the STAR-RIS were designed and predicted.
Moreover, by optimizing the energy splitting factors of the STAR-RIS, we proposed the trade-off design for the performance of sensing and communication.
Simulation results were provided to demonstrate the validity of our proposed STAR-RIS aided ISAC scheme.

%In this paper, we proposed a ISAC scheme for high mobility scenario, where a STAR-RIS is equipped with the moving vehicle.
%Firstly, we designed a transmission structure for ISAC.
%Then, we studied the near-field static channel model inside the vehicle as well as the far-filed time-frequency selective channel model outside the vehicle.
%Moreover, we resorted to the MOMP algorithm to obtain the cascaded channel parameters from the BS to the RSUs.
%To realize the vehicle localization and velocity measurement, we directly gathering the cascaded channel parameters about all the RSUs, instead of decoupling the parameters of each side of the STAR-RIS.
%With the sensing results, the reflection and refraction phase shifts of the STAR-RIS were designed and predicted.
%Furthermore, we proposed the trade-off design for the performance of sensing and communication by allocating the energy splitting factors of the STAR-RIS.
%Simulation results were provided to demonstrate the validity of our proposed ISAC scheme.

\section*{Appendix A\\ Derivation for the Coordinate of the Intersection Point Between the Ellipsoid $\mathscr E_g^{E_g}$ and the Line $\mathscr L_g^{E_g}$ under $\mathcal C^{E_g}$}

The equation set \eqref{equations_for_intersection} can be transformed as
\begin{align}\label{draft_1}
&\bigg(\frac{1}{a_{g}^2} +
\frac{(\mathbf k_{g}^{E_g}[2])^2}{(\mathbf k_{g}^{E_g}[1])^2 b_{g}^2}
+ \frac{(\mathbf k_{g}^{E_g}[3])^2}{(\mathbf k_{g}^{E_g}[1])^2 b_{g}^2}\bigg) x_0^2
\notag\\
&\quad+ 2 \frac{d_{\text{B},g}}{2}
\bigg( \frac{(\mathbf k_{g}^{E_g}[2])^2}{(\mathbf k_{g}^{E_g}[1])^2 b_{g}^2} + \frac{(\mathbf k_{g}^{E_g}[3])^2}{(\mathbf k_{g}^{E_g}[1])^2 b_{g}^2}\bigg) x_0 \notag \\
&\quad+ \bigg(\frac{d_{\text{B},g}}{2} \bigg)^2
\bigg( \frac{(\mathbf k_{g}^{E_g}[2])^2}{(\mathbf k_{g}^{E_g}[1])^2 b_{g}^2} + \frac{(\mathbf k_{g}^{E_g}[3])^2}{(\mathbf k_{g}^{E_g}[1])^2 b_{g}^2}\bigg) \!-\! 1 \!=\! 0.
\end{align}

Its discriminant can be derived as
\begin{align}\label{draft_2}
\Delta \!=& 4 \Big(\frac{1}{a_{g}^2} \!+\!
\frac{(\mathbf k_{g}^{E_g}[2])^2 + \mathbf k_{g}^{E_g}[3])^2}{(b_{g} \mathbf k_{g}^{E_g}[1])^2} \Big)
\notag\\
&-\! 4 \Big(\frac{d_{\text{B},g}}{2} \!\Big)^2
\!\Big(\frac{1}{a_{g}^2} \Big)
\!\Big(\! \frac{(\mathbf k_{g}^{E_g}[2])^2 + \mathbf k_{g}^{E_g}[3])^2}{(b_{g} \mathbf k_{g}^{E_g}[1])^2}\Big)
%\notag \\
%=& 4\frac{(\mathbf k_{g}^{E_g}[1])^2 + (\mathbf k_{g}^{E_g}[2])^2 + (\mathbf k_{g}^{E_g}[3])^2}{(\mathbf k_{g}^{E_g}[1])^2 a_{g}^2}
%\notag\\
\notag\\
=&
\frac{4}{(\mathbf k_{g}^{E_g}[1])^2 a_{g}^2},
\end{align}
where the relationship $\left(\frac{d_{\text{B},g}}{2} \right)^2 = a_{g}^2 -  b_{g}^2$ is utilized.
Since $\mathbf k_{g}^{E_g}[1] \neq 0$, and thus $\Delta > 0$.
Then, the solution of $x_0$ is
\begin{align}\label{draft_3}
x_0
=& \frac{-2\frac{d_{\text{B},g}}{2}\Big( \frac{(\mathbf k_{g}^{E_g}[2])^2}{(\mathbf k_{g}^{E_g}[1])^2 b_{g}^2} + \frac{(\mathbf k_{g}^{E_g}[3])^2}{(\mathbf k_{g}^{E_g}[1])^2 b_{g}^2}\Big)
\pm \sqrt{\Delta}}
{2\Big(\frac{1}{a_{g}^2} +
\frac{(\mathbf k_{g}^{E_g}[2])^2}{(\mathbf k_{g}^{E_g}[1])^2 b_{g}^2} + \frac{(\mathbf k_{g}^{E_g}[3])^2}{(\mathbf k_{g}^{E_g}[1])^2 b_{g}^2}\Big)}
%\notag\\
%=
%\frac{-\frac{d_{\text{B},g}}{2}\left( \frac{(\mathbf k_{g}^{E_g}[2])^2}{(\mathbf k_{g}^{E_g}[1])^2 b_{g}^2} + \frac{(\mathbf k_{g}^{E_g}[3])^2}{(\mathbf k_{g}^{E_g}[1])^2 b_{g}^2}\right)
%\pm \frac{1}{\mathbf k_{g}^{E_g}[1] a_{g}}}
%{\frac{ a_{g}^2 -(\mathbf k_{g}^{E_g}[1])^2 (\frac{d_{\text{B},g}}{2})^2}
%{a_{g}^2 (\mathbf k_{g}^{E_g}[1])^2 b_{g}^2}}
%\notag\\
%=&
%\frac{ \left(-\frac{a_{g}^2 d_{\text{B},g}}{2}\left(1 - (\mathbf k_{g}^{E_g}[1])^2\right)
%\pm b_{g}^2  a_{g} \mathbf k_{g}^{E_g}[1]  \right)}
%{a_{g}^2 -(\mathbf k_{g}^{E_g}[1])^2 (\frac{d_{\text{B},g}}{2})^2}
%\notag \\
\notag\\
=&
%-\frac{d_{\text{B},g}}{2} +
%\frac{b_{g}^2 \mathbf k_{g}^{E_g}[1] \left(\frac{d_{\text{B},g}}{2} \mathbf k_{g}^{E_g}[1] \pm  a_{g}\right)}{a_{g}^2 (1-\mathbf k_{g}^{E_g}[1])^2 ) + b_{g}^2 (\mathbf k_{g}^{E_g}[1])^2 }
%\notag \\
%=&
-\frac{d_{\text{B},g}}{2} -
\frac{b_{g}^2 \mathbf k_{g}^{E_g}[1] }
{\frac{d_{\text{B},g}}{2} \mathbf k_{g}^{E_g}[1] \pm a_{g}}
\end{align}

Define $x_1 = -\frac{d_{\text{B},g}}{2} -
\frac{b_{g}^2 \mathbf k_{g}^{E_g}[1] }
{\frac{d_{\text{B},g}}{2} \mathbf k_{g}^{E_g}[1] + a_{g}}$
and $x_2 = -\frac{d_{\text{B},g}}{2} -
\frac{b_{g}^2 \mathbf k_{g}^{E_g}[1] }
{\frac{d_{\text{B},g}}{2} \mathbf k_{g}^{E_g}[1] - a_{g}}$,
the solutions of two intersection points are
%\begin{align}
%\mathbf p_{I,2}^{E_g} =& (x_1, \frac{\left(x_1 + \frac{d_{\text{B},g}}{2}\right) \mathbf k_{g}^{E_g}[2]}
%{\mathbf k_{g}^{E_g}[1]}, \frac{\left(x_1 + \frac{d_{\text{B},g}}{2}\right) \mathbf k_{g}^{E_g}[3]}
%{\mathbf k_{g}^{E_g}[1]}), \label{xp1_solution}\\
%\mathbf p_{I,2}^{E_g} =& (x_2, \frac{\left(x_2 + \frac{d_{\text{B},g}}{2}\right) \mathbf k_{g}^{E_g}[2]}
%{\mathbf k_{g}^{E_g}[1]}, \frac{\left(x_2 + \frac{d_{\text{B},g}}{2}\right) \mathbf k_{g}^{E_g}[3]}
%{\mathbf k_{g}^{E_g}[1]}). \label{xp2_solution}
%\end{align}
\begin{align}
\mathbf p_{S,1}^{E_g} =& (x_1,
\frac{(x_1 + \frac{d_{\text{B},g}}{2}) \mathbf k_{g}^{E_g}[2]}
{\mathbf k_{g}^{E_g}[1]},
\frac{(x_1 + \frac{d_{\text{B},g}}{2}) \mathbf k_{g}^{E_g}[3]}
{\mathbf k_{g}^{E_g}[1]}), \label{xp_solution_all_1}
\\
\mathbf p_{S,2}^{E_g} =& (x_2,
\frac{(x_2 + \frac{d_{\text{B},g}}{2}) \mathbf k_{g}^{E_g}[2]}
{\mathbf k_{g}^{E_g}[1]},
\frac{(x_2 + \frac{d_{\text{B},g}}{2}) \mathbf k_{g}^{E_g}[3]}
{\mathbf k_{g}^{E_g}[1]}). \label{xp_solution_all_2}
\end{align}

%\begin{align}
%\left\{
%\begin{aligned}
%&x_0 = -\frac{d_{\text{UR}}}{2} -
%\frac{b_{p}^2 \mathbf k_{p}^{E_p}[1]^2 }
%{\frac{d_{\text{UR}}}{2} \mathbf k_{p}^{E_p}[1] \pm a_{p}} \\
%%%%%%%%%%%%%%%%%%%%%%%%%%%%%%%%%%%%%%%%%%%%%%
%&y_0 =
%\frac{\left(x_0 + \frac{d_{\text{UR}}}{2}\right) \mathbf k_{p}^{E_p}[2]}
%{\mathbf k_{p}^{E_p}[1]} \\
%%%%%%%%%%%%%%%%%%%%%%%%%%%%%%%%%%%%%%%%%%%%%%
%&z_0 =
%\frac{\left(x_0 + \frac{d_{\text{UR}}}{2}\right) \mathbf k_{p}^{E_p}[3]}
%{\mathbf k_{p}^{E_p}[1]}
%\end{aligned}
%\right.
%\end{align}

%\linespread{1.44}
\balance

%\balance
%\bibliographystyle{IEEEtran}
%\bibliography{./bibtex/IEEEabrv,./bibtex/ref}

%\balance
%\bibliographystyle{IEEEtran}
%\bibliography{./bibtex/IEEEabrv,./bibtex/ref}
\end{document}